\documentclass[12pt]{article}
\usepackage{amsmath,amssymb,amsfonts,graphicx,cite,colordvi,myfeynarts,soul,stmaryrd,color}

\evensidemargin \oddsidemargin
\newcommand\Code[1]{\ensuremath{\texttt{#1}}}
\newcommand\Var[1]{\ensuremath{\mathit{#1}}}
\newcommand\Vi{\Var{i}}
\newcommand\Vj{\Var{j}}
\newcommand\Vt{\Var{t}}
\newcommand\Vg{\Var{g}}
\newcommand\Vgp{\Var{g'}}
\newcommand\Vh{\Var{h}}
\newcommand\Vhp{\Var{h'}}
\newcommand\Vs{\Var{s}}
\newcommand\Vsp{\Var{s'}}
\newcommand\Vo{\Var{o}}
\newcommand\Vn{\Var{n}}
\newcommand\Vnp{\Var{n'}}
\newcommand\Vc{\Var{c}}
\newcommand\Vcp{\Var{c'}}
\newcommand\Vu{\Var{u}}
\newcommand\Vx{\Var{x}}
\newcommand\opt[2]{\rlap{\raise .7ex\hbox{\small #1}}%
  \lower .9ex\hbox{\vphantom{g}\small #2}}
\newcommand\bbarb{\textrm{[}bar\textrm{]}}
\newcommand\rI{$|$}

\newcommand\TB{t_\beta}
\newcommand\SB{s_\beta}
\newcommand\CB{c_\beta}
\newcommand\CBB{c_{2\beta}}
\newcommand\SA{s_\alpha}
\newcommand\CA{c_\alpha}
\newcommand\CBA{c_{\beta - \alpha}}
\newcommand\SBA{s_{\beta - \alpha}}
\newcommand\CAB{c_{\alpha + \beta}}
\newcommand\SAB{s_{\alpha + \beta}}

\newcommand\LP{\left(}
\newcommand\RP{\right)}
\newcommand\LB{\left[}
\newcommand\RB{\right]}
\newcommand\LV{\left\{}
\newcommand\RV{\right\}}
\newcommand\ri{\mathrm{i}}
\newcommand\re{\mathrm{e}}
\renewcommand\Re{\mathop{\mathrm{Re}}}

\newcommand\ReTilde{\mathop{\widetilde{\mathrm{Re}}}}
\newcommand\ReDiag{\mathop{%
  \raise .5pt\hbox{[}%
  \widetilde{\mathrm{Re}}%
  \raise .5pt\hbox{]}}}
\newcommand\ReOffDiag{\mathop{%
  \raise .5pt\hbox{$\llbracket$}%
  \widetilde{\mathrm{Re}}%
  \raise .5pt\hbox{$\rrbracket$}}}
\newcommand\SE[1]{\Sigma_{#1}}
\newcommand\OS{\mathrm{OS}}
\newcommand\DRbar{\ensuremath{\smash{\overline{\mathrm{DR}}}}}

\newcommand\matr[1]{\mathbf{#1}}
\newcommand\mati[1]{\bigl(#1\bigr)}
\newcommand\unity{\mathrm{1\mskip-4.25mu l}}
\newcommand\ddiv{\bigr|_{\mathrm{div}}}

\newcommand\pslash[1]{\rlap{$\mskip .8mu/$}#1}
\newcommand\boxA{\fbox{\small A}}
\newcommand\boxH{\fbox{\small H}}

\newcommand\SW{s_\mathrm{w}}
\newcommand\CW{c_\mathrm{w}}
\newcommand\MW{M_W}
\newcommand\MZ{M_Z}
\newcommand\MA{M_A}
\newcommand\MHp{M_{H^\pm}}
\newcommand\mf[1]{m_{f_{#1}}}
\newcommand\mb{m_b}
\newcommand\Ab{A_b}

\newcommand\Sf{{\tilde f}}
\newcommand\msf[1]{m_{\Sf_{#1}}}
\newcommand\Snu{{\tilde\nu}}
\newcommand\msnu[1]{m_{\Snu_{#1}}}
\newcommand\Se{\mathrm{\tilde e}}
\newcommand\Fe{\mathrm{e}}
\newcommand\mfe[1]{m_{\Fe_{#1}}}
\newcommand\mse[1]{m_{\Se_{#1}}}
\newcommand\Su{\mathrm{\tilde u}}
\newcommand\Fu{\mathrm{u}}
\newcommand\mfu[1]{m_{\Fu_{#1}}}
\newcommand\msu[1]{m_{\Su_{#1}}}
\newcommand\Sd{\mathrm{\tilde d}}
\newcommand\Fd{\mathrm{d}}
\newcommand\mfd[1]{m_{\Fd_{#1}}}
\newcommand\msd[1]{m_{\Sd_{#1}}}

\newcommand\Stau{{\tilde\tau}}
\newcommand\Stop{{\tilde t}}
\newcommand\Sbot{{\tilde b}}

\newcommand\gl{{\tilde g}}
\newcommand\mgl{m_\gl}
\newcommand\phigl{\varphi_\gl}

\newcommand\cH[1]{\mathcal{H}_{#1}}

\newcommand\tadH{T_H}
\newcommand\tadh{T_h}
\newcommand\tadA{T_A}

\newcommand\dtadH{\delta T_H}
\newcommand\dtadh{\delta T_h}
\newcommand\dtadA{\delta T_A}

\newcommand\dTB{\delta\TB}
\newcommand\dZ[1]{\delta Z_{#1}}
\newcommand\dbZ[1]{\delta\breve Z_{#1}}
\newcommand\dhZ[1]{\delta\hat Z_{#1}}

\newcommand\dZm[1]{\delta\matr{Z}_{#1}}
\newcommand\dbZm[1]{\delta\matr{\breve Z}_{#1}}
\newcommand\dBZm[1]{\delta\matr{\breve{\bar Z}}{}_{#1}}

\newcommand\dMZsq{\delta\MZ^2}
\newcommand\dMWsq{\delta\MW^2}
\newcommand\dMAsq{\delta\MA^2}
\newcommand\dMHpsq{\delta\MHp^2}

\newcommand\ino[1]{\tilde\chi_{#1}}
\newcommand\mino[1]{m_{\ino{#1}}}
\newcommand\chapm[1]{\ino{#1}^\pm}

\newcommand\cham[1]{\ino{#1}^-}
\newcommand\cha{\cham}
\newcommand\mcha[1]{m_{\chapm{#1}}}
\newcommand\neu[1]{\ino{#1}^0}
\newcommand\mneu[1]{m_{\neu{#1}}}

\newcommand\refeq[1]{Eq.~(\ref{#1})}
\newcommand\refeqs[1]{Eqs.~(\ref{#1})}
\newcommand\refta[1]{Table~\ref{#1}}
\newcommand\refse[1]{Sect.~\ref{#1}}
\newcommand\refses[1]{Sects.~\ref{#1}}
\newcommand\citere[1]{Ref.~\cite{#1}}
\newcommand\citeres[1]{Refs.~\cite{#1}}

\newcommand\lbrac{\symbol{123}}
\newcommand\rbrac{\symbol{125}}
\newcommand\Brac[1]{\lbrac#1\rbrac}
\newcommand\uscore{\symbol{95}}
\newcommand\eg{e.g.\ }
\newcommand\ie{i.e.\ }

\DeclareMathOperator{\diag}{diag}

\definecolor{Orange}{named}{Orange}
\definecolor{Purple}{named}{Purple}
\definecolor{Lightblue}{cmyk}{0.9,0.1,0.1,0.3}
\definecolor{dgelborange}{cmyk}{0.,0.3,0.5, 0.}
\definecolor{Lila}{rgb}{0.5,0.,1}

\allowdisplaybreaks
\begin{document}
\thispagestyle{empty}

\begin{flushright}
CERN-PH-TH/2013-168 \\
MPP-2013-185 \\
arXiv:1309.1692 [hep-ph]
\end{flushright}

\vspace{0.5cm}

\begin{center}

\begin{large}
\textbf{The Implementation of the Renormalized Complex MSSM}
\\[2ex]
\textbf{in FeynArts and FormCalc}
\end{large}

\vspace{1cm}

\renewcommand\thefootnote{[ ]}
\footnotetext{former address}
\renewcommand\thefootnote{\fnsymbol{footnote}}

\begin{textsc}
T.~Fritzsche$^{[1]}$%
\footnote{e-mail: Thomas.Fritzsche@de.bosch.com},
T.~Hahn$^1$%
\footnote{e-mail: hahn@feynarts.de},
S.~Heinemeyer$^2$%
\footnote{e-mail: Sven.Heinemeyer@cern.ch}, \\[.2em]
F.~von~der~Pahlen$^2$%
\footnote{e-mail: pahlen@ifca.unican.es, MultiDark Fellow},
H.~Rzehak$^{[3],4}$%
\footnote{e-mail: heidi.rzehak\!\! @cern.ch},
and C.~Schappacher$^{[5]}$%
\footnote{e-mail: schappacher@kabelbw.de}%
\end{textsc}

\vspace*{.7cm}

{\sl
$^1$Max-Planck-Institut f\"ur Physik (Werner-Heisenberg-Institut), \\
F\"ohringer Ring 6,
D--80805 M\"unchen, Germany

\vspace*{0.1cm}

$^2$Instituto de F\'isica de Cantabria (CSIC-UC), Santander, Spain

\vspace*{0.1cm}

$^3$PH-TH, CERN, CH--1211 Gen\`eve 23, Switzerland

\vspace*{0.1cm}

$^4$Albert-Ludwigs-Universit\"at Freiburg, Physikalisches Institut, \\
D--79104 Freiburg, Germany

\vspace*{0.1cm}

$^5$Institut f\"ur Theoretische Physik,
Karlsruhe Institute of Technology, \\
D--76128 Karlsruhe, Germany%
}

\end{center}

\vspace*{0.1cm}

\begin{abstract}
We describe the implementation of the renormalized complex MSSM (cMSSM)
in the diagram generator FeynArts and the calculational tool FormCalc.  
This extension allows to perform UV-finite one-loop calculations of
cMSSM processes almost fully automatically.
The Feynman rules for the cMSSM with counterterms are available as a new 
model file for FeynArts.  Also included are default definitions of the 
renormalization constants; this fixes the renormalization scheme.  
Beyond that all model parameters are generic, \eg we do not impose any
relations to restrict the number of input parameters.
The model file has been tested extensively for several non-trivial 
decays and scattering reactions.  Our renormalization scheme has been 
shown to give stable results over large parts of the cMSSM parameter 
space.

\end{abstract}

\def\thefootnote{\arabic{footnote}}
\setcounter{page}{0}
\setcounter{footnote}{0}

\clearpage
\newpage


\section{Introduction}

One of the main problems of Feynman-diagrammatic computations is the 
enormous growth of the number of Feynman diagrams, not only with the 
loop order, but also with the number of particles in a model.  While 
many precision calculations in the Standard Model (SM) could still be 
performed by hand exactly, the same is very difficult in models like the 
Minimal Supersymmetric Standard Model (MSSM), even more so when 
parameters are allowed to take complex values (cMSSM).  Yet it is highly 
desirable to perform unabridged calculations in the MSSM, too, since 
also this model allows to make precise predictions in terms of a set of 
input parameters.

With the availability of powerful software packages, the basic problem 
of bookkeeping and calculation of the diagrams has been solved for many 
common cases.  Still, it is not entirely trivial to code a model of the 
complexity of the MSSM in such a system, since this has to be done in a 
reasonably general way (\ie not only for special cases of the 
parameters) and many checks have to be performed to test all sectors of 
the model.  Publicly available model files for the MSSM tree-level 
couplings are described in \citeres{MSSMmod,FR,GRACE}.

The present paper documents the implementation of the renormalized cMSSM 
with minimal flavor violation in the FeynArts \cite{feynarts} and 
FormCalc \cite{formcalc} packages.  The counterterms and renormalization 
constants have been tested extensively and show stable results over 
large parts of the cMSSM parameter space.

\refse{sec:diaggen} describes the new MSSM model file, with details of 
the renormalization given in \refse{sec:renorm}.  \refse{sec:usage} 
contains usage information and \refse{sec:tests} lists the calculations 
performed to test the model file.


\section{The MSSMCT model file}
\label{sec:diaggen}

The model file is the source of all physics information in FeynArts.  It 
declares the properties of the fields, their propagators, and their 
couplings. In the model file the generic parameters of the Lagrangian 
are used, not a restricted set of input parameters.

There are two versions of the renormalized MSSM model file in FeynArts, 
both of which follow the conventions (for the MSSM at tree-level) of 
\citere{HaK85,GuH86,hhg} and are based on the existing MSSM model file 
included in FeynArts \cite{MSSMmod}.  The file \Code{MSSMCT.mod} defines 
the complete (electroweak and strong) cMSSM including all counterterms.  
\Code{SQCDCT.mod} contains only the SQCD part, \ie the 
$\alpha_{\text{em}} = 0$ limit, which is extracted from 
\Code{MSSMCT.mod} at load time.

\refta{tab:mssmparticles} gives the names of the fields defined in 
\Code{MSSMCT.mod} and their masses, with index notation in 
\refta{tab:indices}.  The symbols used for the MSSM parameters are 
specified in \refta{tab:modelsyms}.  Pre-defined filters to exclude 
certain groups of particles are listed in \refta{tab:restrict}.  In 
\refta{tab:ct} we give an overview about the newly introduced symbols 
for the renormalization constants.  The ones that appear already in the 
SM part are included for completeness.


\begin{table}[p]
\caption{\label{tab:mssmparticles}The particle content of
\Code{MSSMCT.mod}.}
\begin{center}
\begin{tabular}{|l|l|l|l||l|l|l|l|} \hline
leptons & $f = f^\dagger\!$ & field & mass &
sleptons & $f = f^\dagger\!$ & field & mass \\ \hline
$\nu_g$ & & \Code{F[1,\,\Brac{\Vg}]} & \Code{0} &
  $\tilde\nu_g$ & & \Code{S[11,\,\Brac{\Vg}]} & \Code{MSf} \\
$\ell_g$ & & \Code{F[2,\,\Brac{\Vg}]} & \Code{MLE} &
  $\tilde\ell_g^s$ & & \Code{S[12,\,\Brac{\Vs,\Vg}]} & \Code{MSf} \\
\hline
\multicolumn{6}{c}{} \\[-1ex] \hline
\multicolumn{4}{|l||}{quarks} &
\multicolumn{4}{|l|}{squarks} \\ \hline
$u_g$ & & \Code{F[3,\,\Brac{\Vg,\Vo}]} & \Code{MQU} &
  $\tilde u_g^s$ & & \Code{S[13,\,\Brac{\Vs,\Vg,\Vo}]} & \Code{MSf} \\
$d_g$ & & \Code{F[4,\,\Brac{\Vg,\Vo}]} & \Code{MQD} &
  $\tilde d_g^s$ & & \Code{S[14,\,\Brac{\Vs,\Vg,\Vo}]} & \Code{MSf} \\
\hline
\multicolumn{6}{c}{} \\[-1ex] \hline
\multicolumn{4}{|l||}{gauge bosons} &
\multicolumn{4}{|l|}{neutralinos, charginos} \\ \hline
$\gamma$ & yes & \Code{V[1]} & \Code{0} &
  $\tilde\chi_n^0$ & yes & \Code{F[11,\,\Brac{\Vn}]} & \Code{MNeu} \\
$Z$ & yes & \Code{V[2]} & \Code{MZ} &
  $\tilde\chi_c^-$ & & \Code{F[12,\,\Brac{\Vc}]} & \Code{MCha} \\ 
$W^-$ & & \Code{V[3]} & \Code{MW} & & & & \\ \hline
\multicolumn{6}{c}{} \\[-1ex] \hline
\multicolumn{4}{|l||}{Higgs/Goldstone bosons} &
\multicolumn{4}{|l|}{ghosts} \\ \hline
$h^0$ & yes & \Code{S[1]} & \Code{Mh0} &
	$u_\gamma$ & & \Code{U[1]} & \Code{0} \\
$H^0$ & yes & \Code{S[2]} & \Code{MHH} &
	$u_Z$ & & \Code{U[2]} & \Code{MZ} \\
$A^0$ & yes & \Code{S[3]} & \Code{MA0} &
	$u_+$ & & \Code{U[3]} & \Code{MW} \\
$G^0$ & yes & \Code{S[4]} & \Code{MZ} &
	$u_-$ & & \Code{U[4]} & \Code{MW} \\
$H^-$ & & \Code{S[5]} & \Code{MHp} &
	$u_g$ & & \Code{U[5,\,\Brac{\Vu}]} & \Code{0} \\
$G^-$ & & \Code{S[6]} & \Code{MW} & & & & \\ \hline
\multicolumn{6}{c}{} \\[-1ex] \hline
\multicolumn{4}{|l||}{gluon} &
\multicolumn{4}{|l|}{gluino} \\ \hline
$g$ & yes & \Code{V[5,\,\Brac{\Vu}]} & \Code{0} &
$\tilde g$ & yes & \Code{F[15,\,\Brac{\Vu}]} & \Code{MGl} \\
\hline
\end{tabular}
\end{center}
\end{table}

\begin{table}[p]
\caption{\label{tab:indices}Index labels and ranges used throughout this 
paper.}
\vspace*{-3ex}
\begin{minipage}[t]{.5\hsize}
\begin{alignat*}{2}
\Vg &= \Code{Index[Generation]} &&= 1\dots 3\,, \\
\Vo &= \Code{Index[Colour]}     &&= 1\dots 3\,, \\
\Vu &= \Code{Index[Gluon]}      &&= 1\dots 8\,, \\
\Vs &= \Code{Index[Sfermion]}   &&= 1\dots 2\,, \\
\Vn &= \Code{Index[Neutralino]} &&= 1\dots 4\,, \\
\Vc &= \Code{Index[Chargino]}   &&= 1\dots 2\,. \\
\end{alignat*}
\end{minipage}\begin{minipage}[t]{.5\hsize}
\begin{gather*}
\text{(S)fermions are indexed by\qquad\qquad} \\
t = \begin{cases}
  1 & \text{(s)neutrinos}, \\
  2 & \text{charged (s)leptons}, \\
  3 & \text{up-type (s)quarks}, \\
  4 & \text{down-type (s)quarks}.
\end{cases}
\end{gather*}
\end{minipage}
\end{table}

\begin{table}[p]
\caption{\label{tab:modelsyms}Symbols representing the MSSM parameters 
in \Code{MSSMCT.mod}.  On the Higgs masses see the discussion at the end 
of \refse{sec:higgs}.}
\begin{center}
\begin{tabular}{|l|l|}
\hline
\Code{Mh0}, \Code{MHH}, \Code{MA0}, \Code{MHp} &
	Higgs masses $m_h$, $m_H$, $\MA$, $\MHp$ \\
\Code{Mh0tree}, \Code{MHHtree}, \Code{MA0tree}, \Code{MHptree} &
	tree-level Higgs masses
	\\
\Code{TB}, \Code{CB}, \Code{SB}, \Code{C2B}, \Code{S2B} &
	$\tan\beta$, $\cos\beta$, $\sin\beta$,
	$\cos 2\beta$, $\sin 2\beta$ \\
\Code{CA}, \Code{SA}, \Code{C2A}, \Code{S2A} &
	$\cos\alpha$, $\sin\alpha$,
	$\cos 2\alpha$, $\sin 2\alpha$
	(tree-level $\alpha$) \\
\Code{CAB}, \Code{SAB}, \Code{CBA}, \Code{SBA} &
	$\cos(\alpha + \beta)$, $\sin(\alpha + \beta)$,
	$\cos(\beta - \alpha)$, $\sin(\beta - \alpha)$ \\
\Code{MUE} &
	Higgs-doublet mixing parameter $\mu$ \\ 
\hline
\Code{MGl} &
	gluino mass $\mgl$ \\
\Code{SqrtEGl} &
	root of the gluino phase, $\re^{\ri\phigl/2}$ \\
\Code{MNeu[\Vn]} &
	neutralino masses $m_{\neu{n}}$ \\
\Code{ZNeu[\Vn,\,\Vnp]} &
	neutralino mixing matrix $\matr{N}_{nn'}$ \\
\Code{MCha[\Vc]} &
	chargino masses $\mcha{c}$ \\
\Code{UCha[\Vc,\,\Vcp]}, \Code{VCha[\Vc,\,\Vcp]} &
	chargino mixing matrices $\matr{U}_{cc'}, \matr{V}_{cc'}$ \\ 
\hline
\Code{MSf[\Vs,\,\Vt,\,\Vg]} &
	sfermion masses $m_{\Sf_{t,sg}}$ \\
\Code{USf[\Vt,\,\Vg][\Vs,\,\Vsp]} &
	sfermion mixing matrix $U^{\Sf_{tg}}_{ss'}$ \\
\Code{Af[\Vt,\,\Vg,\,\Vgp] } &
	soft-breaking trilinear $A$-parameters
	$\mati{\matr{A}_{f_t}}_{gg'}$ \\ 
\hline
\Code{MW}, \Code{MZ} &
	gauge-boson masses $\MW$, $\MZ$ \\
\Code{CW}, \Code{SW} &
	$\CW\equiv\cos\theta_{\text{w}} = \MW/\MZ$,
	$\SW\equiv\sin\theta_{\text{w}}$ \\
\Code{EL} &
	electromagnetic coupling constant $e$ \\
\Code{GS} &
	strong coupling constant $g_s$ \\
\hline
\end{tabular}
\end{center}
\end{table}

\begin{table}[p]
\caption{\label{tab:restrict}Particle-exclusion filters defined in 
\Code{MSSMCT.mod}.  Observe that indiscriminate use of these filters
may destroy finiteness of the results.}
\begin{center}
\begin{tabular}{|l|p{.68\hsize}|}
\hline
\Code{NoGeneration1} &
	exclude generation-1 (s)fermions \\
\Code{NoGeneration2} &
	exclude generation-2 (s)fermions \\
\Code{NoGeneration3} &
	exclude generation-3 (s)fermions \\
\hline
\Code{NoElectronHCoupling} &
	\rightskip 0em plus 7em
	exclude all couplings of electrons and any Higgs/Goldstone
	particle \\
\Code{NoLightFHCoupling} &
	\rightskip 0em plus 7em
	exclude all couplings of light fermions (all fermions
	except the top) and any Higgs/Goldstone particle \\ 
\hline
\Code{NoSUSYParticles} &
	\rightskip 0em plus 7em
	exclude particles not present in the SM: sfermions,
	charginos, neutralinos, the Higgs particles
	$H$, $A$, $H^\pm$ \\
\Code{THDMParticles} &
	\rightskip 0em plus 7em
	exclude particles not present in the two-Higgs-doublet
	model: sfermions, charginos, and neutralinos \\ 
\hline
\end{tabular}
\end{center}
\end{table}

\begin{table}[p]
\caption{\label{tab:ct}Renormalization constants (RCs) used in 
\Code{MSSMCT.mod}, where $a$\rI$b$ means `$a$ or $b$' and
\Code{dZ\bbarb} stands for both \Code{dZ} and \Code{dZbar}.}
\vspace*{-2pt}
\begin{center}
\begin{tabular}{|l|l|l|}
\multicolumn{2}{l}{Higgs-boson Sector (\refse{sec:higgs})} &
\multicolumn{1}{l}{Eqs.} \\
\hline
\Code{dZ\bbarb Higgs1[\Vh,\,\Vhp]} &
	Higgs field RCs & \eqref{eq:dZHiggs} \\
\Code{dMHiggs1[\Vh,\,\Vhp]} &
	Higgs mass RCs & \eqref{eq:dMHiggs} \\
\Code{dTh01}, \Code{dTHH1}, \Code{dTA01} &
	Higgs tadpole RCs & \eqref{eq:dTad} \\
\Code{dZH1}, \Code{dZH2},
\Code{dTB1}, \Code{dSB1}, \Code{dCB1} &
	RCs related to $\beta$ & \eqref{eq:dbeta} \\
\hline
\multicolumn{3}{l}{Gauge-boson Sector (\refse{sec:gauge})} \\
\hline
\Code{dMZsq1}, \Code{dMWsq1} &
	gauge-boson mass RCs & \eqref{eq:dMgauge} \\
\Code{dZAA1}, \Code{dZAZ1}, \Code{dZZA1}, \Code{dZZZ1}, \Code{dZ\bbarb W1} &
	gauge-boson field RCs & \eqref{eq:dZgauge} \\
\Code{dSW1}, \Code{dZe1} &
	coupling-constant RCs & \eqref{eq:dZe} \\
\hline
\multicolumn{3}{l}{Chargino/Neutralino Sector (\refse{sec:chaneu})} \\
\hline
\Code{dMCha1[\Vc,\,\Vcp]} &
	chargino mass RCs & \eqref{eq:dMCha} \\
\Code{dMNeu1[\Vn,\,\Vnp]} &
	neutralino mass RCs & \eqref{eq:dMNeu} \\
\Code{dMino11}, \Code{dMino21}, \Code{dMUE1} &
	RCs for $M_1$, $M_2$, $\mu$ & \eqref{eq:dCha} \\
\Code{dZ\bbarb f\opt{L}{R}1[12,\,\Vc,\,\Vcp]} &
	chargino field RCs & \eqref{eq:dZChaNeu} \\
\Code{dZ\bbarb f\opt{L}{R}1[11,\,\Vn,\,\Vnp]} &
	neutralino field RCs & \eqref{eq:dZChaNeu} \\
\hline
\multicolumn{3}{l}{Fermion Sector (\refse{sec:fermion})} \\
\hline
\Code{dMf1[\Vt,\,\Vg]} &
	fermion mass RCs & \eqref{eq:dMf} \\
\Code{dZ\bbarb f\smash{\opt{L}{R}}1[\Vt,\,\Vg,\,\Vgp]} &
	fermion field RCs & \eqref{eq:dZf} \\
\Code{dCKM1[\Vg,\,\Vgp]} &
	CKM-matrix RCs & \eqref{eq:dCKM} \\
\hline
\multicolumn{3}{l}{Squark Sector (\refse{sec:squark})} \\
\hline
\Code{dMSfsq1[\Vs,\,\Vsp,\,3\rI4,\,\Vg]} &
	squark mass RCs & \eqref{eq:dMSq}, \eqref{eq:dMSd} \\
\Code{dAf1[3\rI 4,\,\Vg,\,\Vg]} &
	trilinear squark coupling RCs & \eqref{eq:dAq} \\
\Code{dZ\bbarb Sf\opt{L}{R}1[\Vs,\,\Vsp,\,3\rI4,\,\Vg]} &
	squark field RCs & \eqref{eq:dZSq} \\
\hline
\multicolumn{3}{l}{Slepton Sector (\refse{sec:slepton})} \\
\hline
\Code{dMSfsq1[\Vs,\,\Vsp,\,1\rI 2,\,\Vg]} &
	slepton mass RCs & \eqref{eq:dMSl} \\
\Code{dAf1[2,\,\Vg,\,\Vg]} &
	trilinear slepton coupling RCs & \eqref{eq:dAe} \\
\Code{dZ\bbarb Sf\opt{L}{R}1[\Vs,\,\Vsp,\,1\rI 2,\,\Vg]} &
	slepton field RCs & \eqref{eq:dZSl} \\
\hline
\multicolumn{3}{l}{Gluino Sector (\refse{sec:gluino})} \\
\hline
\Code{dMGl1} &
	gluino mass RC & \eqref{eq:dMGl} \\
\Code{dZ\bbarb Gl\opt{L}{R}1} &
	gluino field RCs & \eqref{eq:dZGl} \\
\hline
\multicolumn{3}{l}{Gluon Sector (\refse{sec:gluon})} \\
\hline
\Code{dZgs1} &
	strong-coupling-constant RC & \eqref{eq:dZgs} \\
\Code{dZGG1} &
	gluon field RCs & \eqref{eq:dZGG} \\
\hline
\end{tabular}
\end{center}
\end{table}

\subsection{Renormalization and Absorptive contributions}

The counterterms have been derived via multiplicative renormalization 
applied to all two-, three- and four-point interactions in the 
Lagrangian.  Special care has been taken to include counterterms that 
appear due to particle mixing for vertices that are zero at the tree 
level, such as the $HZ\gamma$-vertex (which obtains a counterterm 
contribution from the non-vanishing tree-level $HZZ$-vertex and one-loop 
$Z$--$\gamma$ mixing).  Feynman gauge is used throughout the model file.

Absorptive contributions arise from the product of imaginary parts of 
complex couplings in a diagram and imaginary parts of the loop functions 
in wave-function corrections (self-energy insertions on external legs), 
\ie in processes with unstable external particles.  These corrections 
are taken into account via wave-function correction factors $\dbZ{}$, 
not to be confused with the field renormalization constants $\dZ{}$ 
introduced by the multiplicative renormalization procedure.  For the 
off-diagonal wave-function correction factors, this procedure has been 
checked against explicitly including the (renormalized) self-energy type 
corrections of the external legs, and full agreement was found.  The 
corrections from the absorptive parts can be sizable 
\cite{Stop2decay,imim,LHCxC,LHCxN}.  Definitions \Code{dZ\Vx} with 
external masses large enough to develop an absorptive part are 
formulated in \refse{sec:renorm} and in the model file in terms of
\begin{subequations}
\label{eq:rediag}
\begin{alignat}{2}
\Code{ReDiag} &\equiv \ReDiag &
	&\quad \text{for diagonal $\dZ{xx}$}\,, \\
\Code{ReOffDiag} &\equiv \ReOffDiag &
	&\quad \text{for off-diagonal $\dZ{xy}$}\,.
\end{alignat}
\end{subequations}
The default value for both is \Code{Identity}, which means that 
absorptive parts are included ($\Code{dZ\Vx} = \dZ{\Vx} + \dbZ{\Vx}$).  
Redefining \Code{Re}[\Code{Off}]\Code{Diag = ReTilde} switches them off 
($\Code{dZ\Vx} = \dZ{\Vx}$).

\smallskip

$\Code{ReTilde}\equiv\ReTilde$ takes the real part of loop integrals 
only and leaves complex couplings unaffected.  The \Code{dZ\Vx} 
including absorptive parts are in general valid for incoming \Vx\ 
(outgoing $\bar x$) only, for outgoing \Vx\ (incoming $\bar x$) 
\Code{dZbar\Vx} must be used.


\section{Details of the Renormalization}
\label{sec:renorm}

\subsection{Prolegomena}

We use the following short-hands in this section:
\begin{itemize}
\item $c_x\equiv\cos x$, $s_x\equiv\sin x$, $t_x\equiv\tan x$.

\item $\SE{ab}(p^2)\equiv\smash{\lower 23pt\hbox{%
\begin{feynartspicture}(50,50)(1,1)
\FADiagram{}
\FAProp(0,10)(7,10)(0,){Straight}{1}
\FALabel(0,11)[lb]{\small $b$}
\FAProp(20,10)(13,10)(0,){Straight}{-1}
\FALabel(20,11)[rb]{\small $a$}
\FALabel(0,9)[lt]{\scriptsize $p\to$}
\FAVert(10.,10.){-1}
\end{feynartspicture}}}$.

\item $\SE{f}(p^2)\equiv
  \pslash{p}\omega_-\SE{f}^L(p^2) + \pslash{p}\omega_+\SE{f}^R(p^2) +
  \omega_-\SE{f}^{SL}(p^2) + \omega_+\SE{f}^{SR}(p^2)$ for fermions, \\
where $\omega_\pm = \tfrac 12 (\unity\pm\gamma_5)$.

\item $\SE{V}(p^2)\equiv
   -\Bigl(g_{\mu\nu} - \frac{p_\mu p_\nu}{p^2}\Bigr)\SE{V}^T(p^2) -
    \frac{p_\mu p_\nu}{p^2}\SE{V}^L(p^2)$ for vector bosons.

\item $\Sigma'(m^2)\equiv\frac{\partial\Sigma(p^2)}{\partial p^2}
  \big|_{p^2 = m^2}$.

\item $\ReTilde$ takes the real part of loop integrals but leaves 
complex couplings unaffected.

\item $\hat{\Sigma}$ denotes a renormalized self energy.
\end{itemize}


\subsection{The Higgs-boson Sector}
\label{sec:higgs}

The MSSM requires two Higgs doublets which results in five physical 
Higgs bosons, the light and heavy CP-even $h$ and $H$, the CP-odd $A$, 
and the charged Higgs bosons $H^\pm$.  As in the SM, the remaining 
degrees of freedom are taken up by the neutral and charged Goldstone 
bosons $G$ and $G^\pm$.  Taking higher-order corrections in the cMSSM 
into account, the three neutral Higgs bosons mix and give rise to three 
states, $h_1$, $h_2$, $h_3$.

The Higgs sector is fixed at lowest order by choosing a value for $\TB = 
v_2/v_1$, the ratio of the vacuum expectation values of the two Higgs 
doublets, and for the mass $\MA$ of the CP-odd neutral Higgs boson $A$ 
(for real input parameters) or the mass $\MHp$ of the charged Higgs 
boson $H^\pm$ (for complex or real input parameters).  The tree-level 
masses result from
\begin{align}
\label{eq:MHtree}
\begin{pmatrix}
m_h^2 & 0 \\
0 & m_H^2
\end{pmatrix} &= \mathcal{U}_\alpha\begin{pmatrix}
\CB^2\MA^2 + \SB^2\MZ^2 & -\SB\CB (\MA^2 + \MZ^2) \\
-\SB\CB (\MA^2 + \MZ^2) & \SB^2\MA^2 + \CB^2\MZ^2
\end{pmatrix}\mathcal{U}_\alpha^T\,, \\[1ex]
\MHp^2 &= \MA^2 + \MW^2\,.
\end{align}
\refeq{eq:MHtree} also establishes the $\alpha$ used in the following as 
the angle that diagonalizes the tree-level CP-even Higgs-boson mass 
matrix through the rotation $\mathcal{U}_\alpha = \LP\begin{smallmatrix} 
\CA & -\SA \\ \SA & \CA \end{smallmatrix}\RP$.

Our multiplicative renormalization procedure shifts masses and couplings 
but not mixing matrices, with the subtle effect that, while the 
$\beta$'s in the mass matrices contribute to the counterterms as 
expected, the ones do not that enter through the $\mathcal{U}_\beta = 
\LP\begin{smallmatrix} \CB & -\SB \\ \SB & \CB \end{smallmatrix}\RP$ in
\begin{align}
\begin{pmatrix}
M_X^2 & 0 \\
0 & 0
\end{pmatrix} &= \mathcal{U}_\beta\begin{pmatrix}
\CB^2 M_X^2 & -\SB\CB M_X^2 \\
-\SB\CB M_X^2 & \SB^2 M_X^2
\end{pmatrix}\mathcal{U}_\beta^T\,,
\qquad
X = A, H^\pm\,.
\end{align}
More details on the renormalization of this sector can be found in
\citeres{mhcMSSMlong,Stop2decay}.

The field renormalization constants are as follows.  Note that those of 
the CP-violating self-energies are zero due to the Higgs potential being 
CP-conserving in lowest order.
\begin{subequations}
\label{eq:dZHiggs}
\begin{alignat}{2}
\Code{dZHiggs1[1,\,1]} &\equiv \dZ{hh} &
	&= \SA^2\dZ{\cH1} + \CA^2\dZ{\cH2}\,, \\
\Code{dZHiggs1[2,\,2]} &\equiv \dZ{HH} &
	&= \CA^2\dZ{\cH1} + \SA^2\dZ{\cH2}\,, \\
\Code{dZHiggs1[3,\,3]} &\equiv \dZ{AA} &
	&= \SB^2\dZ{\cH1} + \CB^2\dZ{\cH2}\,, \\
\Code{dZHiggs1[4,\,4]} &\equiv \dZ{GG} &
	&= \CB^2\dZ{\cH1} + \SB^2\dZ{\cH2}\,, \\[1ex]
\Code{dZHiggs1[1,\,2]} &\equiv \dZ{hH} &
	&= \SA\CA (\dZ{\cH2} - \dZ{\cH1})\,, \\[1ex]
\Code{dZHiggs1[1,\,3]} &\equiv \dZ{hA} &
	&= 0\,, \\
\Code{dZHiggs1[2,\,3]} &\equiv \dZ{HA} &
	&= 0\,, \\[1ex]
\Code{dZHiggs1[1,\,4]} &\equiv \dZ{hG} &
	&= 0\,, \\
\Code{dZHiggs1[2,\,4]} &\equiv \dZ{HG} &
	&= 0\,, \\
\Code{dZHiggs1[3,\,4]} &\equiv \dZ{AG} &
	&= \SB\CB (\dZ{\cH2} - \dZ{\cH1})\,, \\[1ex]
\label{eq:dZHpHm}
\Code{dZHiggs1[5,\,5]} &\equiv \dZ{H^-H^+} + \dhZ{H^-H^+} + \dbZ{H^-H^+} &
	&= -\ReDiag\Sigma'_{H^-}(\MHp^2)\,, \\
\Code{dZbarHiggs1[5,\,5]}
	&= \rlap{\Code{dZHiggs1[5,\,5]}\,,} \\
\Code{dZHiggs1[6,\,6]} &\equiv \dZ{G^-G^+} &
	&= \dZ{GG}\,, \\[1ex]
\Code{dZHiggs1[5,\,6]} &\equiv \dZ{H^-G^+} &
	&= \dZ{AG}\,, \\[2ex]
\Code{dZHiggs1[\Vj,\,\Vi]}
	&= \rlap{\Code{dZHiggs1[\Vi,\,\Vj]}\,.}
\end{alignat}
\end{subequations}

Of the three contributions in \refeq{eq:dZHpHm}, $\dZ{H^-H^+}$ $(= 
\dZ{AA})$ contains only the UV-divergent part and is sufficient to yield 
UV-finite results, yet $\dhZ{H^-H^+}$ is needed to render the results 
IR-finite and ensure on-shell properties of an external charged Higgs 
boson.  Corrections to the charged-Higgs-boson propagator give rise to 
the extra factor
\begin{equation}
\hat Z_{H^-H^+}\equiv 1 + \dhZ{H^-H^+}
	= \LP 1 + \ReTilde\hat\Sigma'_{H^-}(\MHp^2)\RP^{-1}
\end{equation}
which, when expanded to one-loop order, leads to $\dhZ{H^-H^+} =
-\ReTilde\Sigma'_{H^-}(\MHp^2) - \dZ{H^-H^+}$ and thus to 
\refeq{eq:dZHpHm}.

The variable \Code{\$MHpInput} (set \emph{before} model initialization) 
chooses the input mass:
\begin{subequations}
\label{eq:MHpInput}
\begin{alignat}{2}
\Code{\$MHpInput}~ &\Code{= False} &\qquad
	&\text{input mass $\MA$}~(\text{case \boxA\ below})\,, \\
\Code{\$MHpInput}~ &\Code{= True} &
	&\text{input mass $\MHp$}~(\text{default: case \boxH\ below})\,.
\end{alignat}
\end{subequations}
With $C = e/(2\MZ\SW\CW)$ the mass counterterms read 
\cite{mhcMSSMlong}\footnote{%
  In \citere{mhcMSSMlong} a slightly different renormalization 
  prescription for $\TB$ had been introduced, $\TB \to \TB (1 +
  \dTB^{\!\hbox{\tiny\cite{mhcMSSMlong}}})$, such that $\dTB = \TB
  \dTB^{\!\hbox{\tiny \cite{mhcMSSMlong}}}$.%
}:
\begin{subequations}
\label{eq:dMHiggs}
\begin{alignat}{2}
\Code{dMHiggs1[1,\,1]} &\equiv \delta m_h^2 &
	&= C\bigl(\CBA\SBA^2\dtadH - \SBA (1 + \CBA^2)\dtadh\bigr) + {} \\
	&&&\qquad 2\CB^2 (\SAB\CAB\MZ^2 - \SBA\CBA\MA^2) \dTB + {} \notag \\
	&&&\qquad \CBA^2\delta\MA^2 + \SAB^2\delta\MZ^2\,, \notag \\
\Code{dMHiggs1[2,\,2]} &\equiv \delta m_H^2 &
	&= -C\bigl(\CBA (1 + \SBA^2)\dtadH - \SBA\CBA^2\dtadh\bigr) - {} \\
	&&&\qquad 2\CB^2 (\SAB\CAB\MZ^2 - \SBA\CBA\MA^2) \dTB + {} \notag \\
	&&&\qquad \SBA^2\delta\MA^2 + \CAB^2\delta\MZ^2\,, \notag \\
\Code{dMHiggs1[3,\,3]} &\equiv \delta\MA^2 &
	&= \begin{cases}
	\ReTilde\SE{A}(\MA^2)	& \boxA\,, \\
	\dMHpsq - \dMWsq	& \boxH\,,
	\end{cases} \\
\Code{dMHiggs1[4,\,4]} &\equiv \delta m_G^2 &
	&= -C \bigl(\CBA\dtadH + \SBA\dtadh\bigr)\,, \\[1ex]
\Code{dMHiggs1[1,\,2]} &\equiv \delta m_{hH}^2 &
	&= -C \bigl(\SBA^3\dtadH + \CBA^3\dtadh\bigr) - {} \\
	&&&\qquad \CB^2 \bigl((\CAB^2 - \SAB^2)\MZ^2 +
	                      (\CBA^2 - \SBA^2)\MA^2\bigr) \dTB - {} \notag \\
	&&&\qquad \SBA\CBA\delta\MA^2 - \SAB\CAB\delta\MZ^2\,, \notag \\[1ex]
\Code{dMHiggs1[1,\,3]} &\equiv \delta m_{hA}^2 &
	&= -C\SBA\dtadA\,, \\
\Code{dMHiggs1[2,\,3]} &\equiv \delta m_{HA}^2 &
	&= -C\CBA\dtadA\,, \\[1ex]
\Code{dMHiggs1[1,\,4]} &\equiv \delta m_{hG}^2 &
	&= -\delta m_{HA}^2\,, \\
\Code{dMHiggs1[2,\,4]} &\equiv \delta m_{HG}^2 &
	&= \delta m_{hA}^2\,, \\
\Code{dMHiggs1[3,\,4]} &\equiv \delta m_{AG}^2 &
	&= C \bigl(\SBA\dtadH - \CBA\dtadh\bigr) - \CB^2\dTB\begin{cases}
	\MA^2			& \boxA\,, \\
	(\MHp^2 - \MW^2)	& \boxH\,,
	\end{cases} \\[-1ex]
\Code{dMHiggs1[\Vj,\,\Vi]}
	&= \rlap{$\Code{dMHiggs1[\Vi,\,\Vj]}$\qquad
	   $i, j = 1\dots 4$\,,} \\[1ex]
\Code{dMHiggs1[5,\,5]} &\equiv \dMHpsq &
	&= \begin{cases}
	\dMAsq + \dMWsq			& \boxA\,, \\
	\ReTilde\SE{H^-}(\MHp^2)	& \boxH\,,
	\end{cases} \\
\Code{dMHiggs1[6,\,6]} &\equiv \delta m_{G^\pm}^2 &
	&= \delta m_G^2\,, \\[1ex]
\Code{dMHiggs1[5,\,6]} &\equiv \delta m_{H^-G^+}^2 &
	&= C \bigl(\SBA\dtadH - \CBA\dtadh + \ri\,\dtadA\bigr) -
              \CB^2\MHp^2\dTB\,, \\
\Code{dMHiggs1[6,\,5]}
	&= \rlap{$\Code{dMHiggs1[5,\,6]}^*$\,.}
\end{alignat}
\end{subequations}
At tree-level the CP-even Higgs fields do not mix with the $Z$ boson 
and hence there are no counterterm contributions to this mixing at 
one-loop level.

\medskip

The tadpole parameters are renormalized such that the complete one-loop
tadpole contributions vanish, leading to
\begin{subequations}
\label{eq:dTad}
\begin{alignat}{2}
\Code{dTh01} &\equiv \dtadh &&= -\tadh^{(1)}\,, \\
\Code{dTHH1} &\equiv \dtadH &&= -\tadH^{(1)}\,, \\
\Code{dTA01} &\equiv \dtadA &&= -\tadA^{(1)} 
\end{alignat}
\end{subequations}
where $T_\phi^{(1)}$ contains all irreducible one-loop tadpole diagrams 
of field $\phi = h, H, A$.

\DRbar\ renormalization of $\dZ{\cH1}$, $\dZ{\cH2}$, and $\dTB$ avoids 
large (and unphysical) higher-order corrections in the Higgs-mass 
calculations \cite{tbren1}.  It has been shown to yield stable numerical 
results \cite{tbren1,tbren2,mhiggsf1lA} and is also gauge-independent at 
the one-loop level within the class of $R_\xi$ gauges \cite{tbren2}.  
Furthermore, there is no obvious relation of $\TB$ to a specific 
physical observable that would favor a particular on-shell definition 
and the divergent part of $\dTB$ can be expressed by the UV-divergent 
parts of the field renormalization constants \cite{tbren3}.
\begin{subequations}
\label{eq:dbeta}
\begin{alignat}{2}
\label{eq:dZH1}
\Code{dZH1} &\equiv \dZ{\cH1} &
	&= -\Re\Sigma'_H(0)\bigr|_{\alpha = 0, \mathrm{div}}\,, \\
\label{eq:dZH2}
\Code{dZH2} &\equiv \dZ{\cH2} &
	&= -\Re\Sigma'_h(0)\bigr|_{\alpha = 0, \mathrm{div}}\,, \\
\Code{dTB1} &\equiv \dTB &
	&= \tfrac 12\TB (\dZ{\cH2} - \dZ{\cH1})\,, \\
\Code{dSB1} &\equiv \delta\SB &
	&= \CB^3\, \dTB\,, \\  
\Code{dCB1} &\equiv \delta\CB &
	&= -\SB\, \CB^2\, \dTB\,,
\end{alignat}
\end{subequations}
\ie the counterterms in \refeqs{eq:dbeta} contribute only UV-divergent
parts\footnote{%
  The divergences in \refeqs{eq:dZH1}, \eqref{eq:dZH2} are
  momentum-independent.
}, and the finite result depends on the renormalization scale $\mu_R$
(in FormCalc: \Code{MUDIM}).

The Higgs-boson field renormalization constants are necessary to render 
the one-loop calculations of partial decay widths with external Higgs 
bosons UV-finite.  The \DRbar\ scheme for the field renormalization 
constants is used in the calculation of the Higgs masses within 
FeynHiggs~\cite{feynhiggs,mhiggslong,mhiggsAEC,mhcMSSMlong} (see 
\refse{sec:numeval} on how to use FeynHiggs with FormCalc) in order to 
avoid the possible occurrence of unphysical threshold effects.

When composing a vertex $\Gamma_{h_i}$ ($i = 1,2,3$) from the 
corresponding tree-level amplitudes $\Gamma_h$, $\Gamma_H$, and 
$\Gamma_A$, another set of finite $Z$-factors is needed to ensure
correct on-shell properties of the external Higgs boson $h_i$ 
\cite{mhcMSSMlong},
\begin{equation}
\label{eq:zfactors123}
\Gamma_{h_i} = \hat Z_{i1}\Gamma_h + \hat Z_{i2}\Gamma_H +
  \hat Z_{i3}\Gamma_A + \ldots\,,
\end{equation}
where the ellipsis represents contributions from the mixing with the 
Goldstone and $Z$ boson, which have to be taken into account explicitly.  
The $Z$-factor matrix $\hat Z_{ij}\equiv\Code{ZHiggs[\Vi,\,\Vj]}$ is not 
in general unitary.  Its lower $3\times 3$ part is computed by FeynHiggs 
and application at the amplitude level automatically takes any 
absorptive contribution into account.  Technically this is most easily 
accomplished using the FeynArts add-on model file \Code{HMix.mod} 
\cite{SUSY07} which mixes $h = \Code{S[1]}$, $H = \Code{S[2]}$, and
$A = \Code{S[3]}$ into two variants of the loop-corrected states $h_i$,
\begin{subequations}
\label{eq:hmix}
\begin{alignat}{2}
\Code{S[0,\,\Brac{\Vi}]} &=
	\sum_{j = 1}^3 \Code{UHiggs[\Vi,\,\Vj]}~\Code{S[\Vj]}\,, &
	&\qquad\begin{minipage}[t]{.4\hsize}
	\raggedright
	with unitary \Code{UHiggs} (no absorptive \\
	contrib.), for use on internal lines,
	\end{minipage} \\
\Code{S[10,\,\Brac{\Vi}]} &=
	\sum_{j = 1}^3 \Code{ZHiggs[\Vi,\,\Vj]}~\Code{S[\Vj]}\,, &
	&\qquad\text{inserted only on external lines.}
\end{alignat}
\end{subequations}
For an external charged Higgs boson a similar factor 
$\sqrt{\smash[b]{\hat Z_{H^-H^+}}} = 1 + \frac 12\dhZ{H^-H^+}$ is 
necessary; this is already included in \refeq{eq:dZHpHm}.  Mixing with 
the Goldstone and the $W$ boson must be computed explicitly, however, as 
in the neutral case.

Finally, a note on the Higgs masses used in the model file.  
Higher-order corrections are phenomenologically very important in the 
Higgs sector, yet the use of loop-corrected masses, besides mixing 
orders in perturbation theory, entails a certain risk of upsetting the 
relations necessary for the proper cancellation of UV and IR 
divergences.  For instance, the masses on the Higgs propagators should 
be consistent with the mixing angle $\alpha$ parameterizing the 
vertices, but this is not easy to achieve in practice since at loop 
level there is mixing between all three states $h$, $H$, $A$, which is 
not expressible through a single angle $\alpha$.

We opted therefore to formulate the vertices with tree-level $\alpha$ 
and insert tree-level masses \Code{Mh0tree}, \Code{MHHtree}, 
\Code{MA0tree}, \Code{MHptree} on loop propagators, where the chance of 
violating supersymmetric relations (and double-counting higher-order 
contributions) is highest, but use loop-corrected masses \Code{Mh0}, 
\Code{MHH}, \Code{MA0}, \Code{MHp} on all other propagators.  At the 
level of the Feynman rules it is not possible to generally avoid 
incomplete cancellations due to a mismatch between tree-level and 
loop-corrected masses, though there are typically process-specific 
solutions (see \eg \citeres{Stop2decay,HWW}).  Our recommendation is to 
test UV and IR finiteness with loop-corrected masses and revert to 
tree-level masses as far as necessary.


\subsection{The Gauge-boson Sector}
\label{sec:gauge}

For the SM gauge bosons we impose the usual set of on-shell renormalization 
conditions and obtain \cite{denner,Stop2decay}
\begin{subequations}
\label{eq:dMgauge}
\begin{alignat}{2}
\Code{dMZsq1} &\equiv \dMZsq &
	&= \ReTilde\SE{Z}^T(\MZ^2)\,, \\
\Code{dMWsq1} &\equiv \dMWsq &
	&= \ReTilde\SE{W}^T(\MW^2)\,,
\end{alignat}
\end{subequations}
\vspace*{-4ex}
\begin{subequations}
\label{eq:dZgauge}
\begin{alignat}{2}
\Code{dZAA1} &\equiv \dZ{\gamma\gamma} &
	&= -\ReTilde\SE{\gamma}^{\prime T}(0)\,, \\
\Code{dZAZ1} &\equiv \dZ{\gamma Z} &
	&= -\frac 2{\MZ^2}\ReTilde\SE{\gamma Z}^T(\MZ^2)\,, \\
\Code{dZZA1} &\equiv \dZ{Z \gamma} &
	&= \frac 2{\MZ^2}\ReTilde\SE{\gamma Z}^T(0)\,, \\
\Code{dZZZ1} &\equiv \dZ{ZZ} + \dbZ{ZZ} &
	&= -\ReDiag\SE{Z}^{\prime T}(\MZ^2)\,, \\
\Code{dZW1} &\equiv \dZ{WW} + \dbZ{WW} &
	&= -\ReDiag\SE{W}^{\prime T}(\MW^2)\,, \\
\Code{dZbarW1}
	&= \rlap{\Code{dZW1}\,.}
\end{alignat}
\end{subequations}

For convenience we define the (dependent) coupling-constant counterterms
\begin{subequations}
\label{eq:dZe}
\begin{alignat}{2}
\Code{dSW1} &\equiv \delta\SW &
	&= \frac 12\frac{\CW^2}{\SW}\LP
	     \frac{\delta\MZ^2}{\MZ^2} - \frac{\delta\MW^2}{\MW^2} \RP, \\
\Code{dZe1} &\equiv \dZ{e} &
	&= \frac 12 \LP \frac{\SW}{\CW}\dZ{Z\gamma} - \dZ{\gamma\gamma}\RP.
\end{alignat}
\end{subequations}

Renormalization constants for the ghost fields are not defined as this 
is necessary only for two-loop calculations.


\subsection{The Chargino/Neutralino Sector}
\label{sec:chaneu}

The chargino/neutralino sector contains two soft-SUSY-breaking gaugino 
mass parameters, $M_1$ for the bino and $M_2$ for the wino field, and 
the Higgs superfield mixing parameter $\mu$, all of which are in general 
complex.\footnote{%
  Our $M_2$ is a complex parameter even though $M_2$ can be chosen real
  without loss of generality.  This is possible since not all phases 
  of the MSSM Lagrangian are physical and there is a certain freedom
  of choice.}
Details on the renormalization can be found in 
\citeres{Stop2decay,LHCxC,LHCxN}.

The chargino masses $\mcha{1,2}$ are obtained from the singular value 
decomposition
\begin{align}
\matr{M}_{\cham{}}\equiv\begin{pmatrix}
  \mcha1 & 0 \\ 
  0 & \mcha2
\end{pmatrix} 
= \matr{V}^*\matr{X}^T\matr{U}^\dagger
\quad\text{with}\quad
\matr{X} = \begin{pmatrix}
  M_2 & \sqrt 2\,\MW\SB \\
  \sqrt 2\,\MW\CB & \mu
\end{pmatrix}.
\end{align}
The neutralino mass matrix
\begin{equation}
\matr{Y} = \begin{pmatrix}
  M_1        & 0          & -\MZ\SW\CB & \MZ\SW\SB  \\
  0          & M_2        & \MZ\CW\CB  & -\MZ\CW\SB \\
  -\MZ\SW\CB & \MZ\CW\CB  & 0          & -\mu       \\
  \MZ\SW\SB  & -\MZ\CW\SB & -\mu       & 0
\end{pmatrix}
\end{equation}
is symmetric and the masses are determined from a Takagi factorization 
\cite{Takagi}
\begin{align}
\matr{M}_{\neu{}} = \matr{N}^*\matr{Y}\,\matr{N}^\dagger =
\diag(m_{\neu{1}}, m_{\neu{2}}, m_{\neu{3}}, m_{\neu{4}})\,.
\end{align}

Parameters and fields are renormalized multiplicatively, following the 
prescription of \citere{dissTF}.  The parameter counterterms are complex 
and thus two renormalization conditions must be specified for each.  The 
transformation matrices $\matr{U}$ and $\matr{V}$ do not obtain 
counterterms, only the mass matrices pick up shifts $\matr{X}\to 
\matr{X} + \delta\matr{X}$ and $\matr{Y}\to \matr{Y} + \delta\matr{Y}$, 
given by
\begin{align}
\label{eq:deltaX}
\delta\matr{X} &= \begin{pmatrix}
  \delta M_2 & \sqrt 2\,\delta(\MW\SB) \\
  \sqrt 2\,\delta(\MW\CB) & \delta\mu
\end{pmatrix}, \\
\delta\matr{Y} &= \begin{pmatrix} 
  \delta M_1         & 0                  & -\delta(\MZ\SW\CB) & \delta(\MZ\SW\SB)  \\
  0                  & \delta M_2         & \delta(\MZ\CW\CB)  & -\delta(\MZ\CW\SB) \\
  -\delta(\MZ\SW\CB) & \delta(\MZ\CW\CB)  & 0                  & -\delta\mu         \\
  \delta(\MZ\SW\SB)  & -\delta(\MZ\CW\SB) & -\delta\mu         &  0
\end{pmatrix}.
\end{align}
This leads to the mass shifts
\begin{alignat}{2}
\matr{M}_{\cham{}}
	&\to \matr{M}_{\cham{}} + \delta\matr{M}_{\cham{}} &
	&= \matr{M}_{\cham{}} +
	   \matr{V}^*\delta\matr{X}^T\matr{U}^\dagger, \\
\matr{M}_{\neu{}}
	&\to \matr{M}_{\neu{}} + \delta\matr{M}_{\neu{}} &
	&= \matr{M}_{\neu{}} +
	   \matr{N}^*\delta\matr{Y}\,\matr{N}^\dagger
\label{eq:Mneu}
\end{alignat}
from which the mass counterterms can be read off as
\begin{align}
\label{eq:dMCha}
\Code{dMCha1[\Vc,\,\Vcp]} &\equiv \LP\delta\matr{M}_{\cham{}}\RP_{cc'}
  = \LP\matr{V}^*\delta\matr{X}^T\matr{U}^\dagger\RP_{cc'}, \\
\label{eq:dMNeu}
\Code{dMNeu1[\Vn,\,\Vnp]} &\equiv \LP\delta\matr{M}_{\neu{}}\RP_{nn'}  
  = \LP\matr{N}^*\delta\matr{Y}\,\matr{N}^\dagger\RP_{nn'}.
\end{align}
The field renormalization constants are \cite{dissTF,Stop2decay}
($c\neq c'$, $n\neq n'$):
\begin{subequations}
\label{eq:dZChaNeu}
\begin{alignat}{2}
\Code{dZfL1[12,\,\Vc,\,\Vc]}
	&\equiv \mati{\dZm{\cha{}}^L + \dbZm{\cha{}}^L}_{cc} &
	&= \mati{\sigma_{cc}^L(\cha{}) + \tau_{cc}(\cha{})}_{cc}\,, \\
\Code{dZbarfL1[12,\,\Vc,\,\Vc]}
	&\equiv \mati{\dZm{\cha{}}^L + \dBZm{\cha{}}^L}_{cc} &
	&= \mati{\sigma_{cc}^L(\cha{}) - \tau_{cc}(\cha{})}_{cc}\,, \\
\Code{dZfR1[12,\,\Vc,\,\Vc]}
	&\equiv \mati{\dZm{\cha{}}^R + \dbZm{\cha{}}^R}_{cc} &
	&= \mati{\sigma_{cc}^R(\cha{}) - \tau_{cc}(\cha{})}_{cc}\,, \\
\Code{dZbarfR1[12,\,\Vc,\,\Vc]}
	&\equiv \mati{\dZm{\cha{}}^R + \dBZm{\cha{}}^R}_{cc} &
	&= \mati{\sigma_{cc}^R(\cha{}) + \tau_{cc}(\cha{})}_{cc}\,,
\\[1ex]
\Code{dZfL1[11,\,\Vn,\,\Vn]}
	&\equiv \mati{\dZm{\neu{}}^L + \dbZm{\neu{}}^L}_{nn} &
	&= \mati{\sigma_{nn}^L(\neu{}) + \tau_{nn}(\neu{})}_{nn}\,, \\
\Code{dZbarfL1[11,\,\Vn,\,\Vn]}
	&= \rlap{\Code{dZfR1[11,\,\Vn,\,\Vn]}\,,} \\
\Code{dZfR1[11,\,\Vn,\,\Vn]}
	&\equiv \mati{\dZm{\neu{}}^R + \dbZm{\neu{}}^R}_{nn} &
	&= \mati{\sigma_{nn}^R(\neu{}) - \tau_{nn}(\neu{})}_{nn}\,, \\
\Code{dZbarfR1[11,\,\Vn,\,\Vn]}
	&= \rlap{\Code{dZfL1[11,\,\Vn,\,\Vn]}\,,}
\end{alignat}
\vspace*{-3.5ex}
\begin{align*}
\text{where}\quad
\sigma_{ii}^X(\ino{}) &= -\ReDiag\Bigl[
	\SE{\ino{}}^X(\mino{i}^2) +
	\!\begin{aligned}[t]
	&\mino{i}^2 \LP\SE{\ino{}}^{\prime L}(\mino{i}^2) +
	               \SE{\ino{}}^{\prime R}(\mino{i}^2)\RP + {} \\
	&\mino{i} \LP\SE{\ino{}}^{\prime SL}(\mino{i}^2) +
	             \SE{\ino{}}^{\prime SR}(\mino{i}^2)\RP
	\Big],
	\end{aligned} \\
\tau_{ii}(\ino{}) &= \frac 1{2\mino{i}}\LP
	\ReDiag\LB\SE{\ino{}}^{SL}(\mino{i}^2) -
	          \SE{\ino{}}^{SR}(\mino{i}^2)\RB -
	\delta\matr{M}_{\ino{}} +
	\delta\matr{M}_{\ino{}}^\dagger\RP.
\end{align*}

\begin{alignat}{2}
\Code{dZfL1[12,\,\Vc,\,\Vcp]}
	&\equiv \mati{\dZm{\cha{}}^L + \dbZm{\cha{}}^L}_{cc'} &
	&= \mati{\sigma_{cc'}^{L,R,SR,SL}(\cha{}) - 
	         \tau_{cc'}(\cha{})}_{cc'}\,, \\
\Code{dZbarfL1[12,\,\Vc,\,\Vcp]}
	&\equiv \mati{\dZm{\cha{}}^L + \dBZm{\cha{}}^L}_{cc'} &
	&= \mati{\sigma_{c'c}^{L,R,SL,SR}(\cha{}) -
	         \tau_{c'c}^\dagger(\cha{})}_{cc'}\,, \\
\Code{dZfR1[12,\,\Vc,\,\Vcp]}
	&\equiv \mati{\dZm{\cha{}}^R + \dbZm{\cha{}}^R}_{cc'} &
	&= \mati{\sigma_{cc'}^{R,L,SL,SR}(\cha{}) -
	         \tau_{cc'}^\dagger(\cha{})}_{cc'}\,, \\
\Code{dZbarfR1[12,\,\Vc,\,\Vcp]}
	&\equiv \mati{\dZm{\cha{}}^R + \dBZm{\cha{}}^R}_{cc'} &
	&= \mati{\sigma_{c'c}^{R,L,SR,SL}(\cha{}) -
	         \tau_{c'c}(\cha{})}_{cc'}\,,
\\[1ex]
\Code{dZfL1[11,\,\Vn,\,\Vnp]}
	&\equiv \mati{\dZm{\neu{}}^L + \dbZm{\neu{}}^L}_{nn'} &
	&= \mati{\sigma_{nn'}^{L,R,SR,SL}(\neu{}) -
	         \tau_{nn'}(\neu{})}_{nn'}\,, \\
\Code{dZbarfL1[11,\,\Vn,\,\Vnp]}
	&= \rlap{\Code{dZfR1[11,\,\Vnp,\,\Vn]}\,,} \\
\Code{dZfR1[11,\,\Vn,\,\Vnp]}
	&\equiv \mati{\dZm{\neu{}}^R + \dbZm{\neu{}}^R}_{nn'} &
	&= \mati{\sigma_{nn'}^{R,L,SL,SR}(\neu{}) -
	         \tau_{nn'}^\dagger(\neu{})}_{nn'}\,, \\
\Code{dZbarfR1[11,\,\Vn,\,\Vnp]}
	&= \rlap{\Code{dZfL1[11,\,\Vnp,\,\Vn]}\,,}
\end{alignat}
\vspace*{-3.5ex}
\begin{align*}
\text{where}\quad
\sigma_{ij}^{X,Y,SX,SY}(\ino{})
	&= \frac 2{\mino{i}^2 - \mino{j}^2} \ReOffDiag\Bigl[
	\mino{j}
	\!\begin{aligned}[t]
	\bigl( &\mino{j}\SE{\ino{}}^X(\mino{j}^2) +
	        \mino{i}\SE{\ino{}}^Y(\mino{j}^2)\bigr) + {} \\
	&\mino{j} \SE{\ino{}}^{SX}(\mino{j}^2) +
	 \mino{i} \SE{\ino{}}^{SY}(\mino{j}^2)\Bigr],
	\end{aligned} \\
\tau_{ij}(\ino{}) &= \frac 2{\mino{i}^2 - \mino{j}^2}
	\ReOffDiag\Bigl[
	\mino{i}\delta\matr{M}_{\ino{}} +
	\mino{j}\delta\matr{M}_{\ino{}}^\dagger\Bigr].
\end{align*}
\end{subequations}
Note again the inclusion of absorptive parts through the $\dbZm{}$.
More detailed information on this for the chargino/neutralino sector can
be found in \citeres{Stop2decay,dissAF,LHCxC,LHCxN}.

Rather than renormalizing the three complex input parameters $M_1$, 
$M_2$, and $\mu$ directly, we impose on-shell conditions for either two 
charginos and one neutralino (CCN) or one chargino and two neutralinos 
(CNN), and from them work out the counterterms $\delta M_1$, 
$\delta M_2$, and $\delta\mu$.


\subsubsection{CCN Schemes}

Inverting \refeqs{eq:deltaX}--\eqref{eq:Mneu} for neutralino $\neu{n}$ 
on-shell yields the following shifts of the breaking parameters 
\cite{dissTF,diplTF}
\begin{subequations}
\label{eq:dCha}
\begin{alignat}{2}
\Code{dMino11} &\equiv \delta M_1 &
	&= \frac 1{N_{n1}^{*2}} \LV \delta\mneu{n}^\OS + \delta N_n -
	  N_{n2}^{*2}\, \delta M_2 +
	  2 N_{n3}^* N_{n4}^*\, \delta\mu \RV, \\
\Code{dMino21} &\equiv \delta M_2 &
	&= \frac 1{u_o v_o - u_d v_d}\biggl\{
	  U_{12}^* V_{12}^*\, \delta\mcha{2}^\OS -
	  U_{22}^* V_{22}^*\, \delta\mcha{1}^\OS + {} \\
&&&\qquad \sqrt 2\,\Bigl(
	    \CB (u_d - u_o) V_{12}^* V_{22}^* +
	    \SB (v_o - v_d) U_{12}^* U_{22}^*
	  \Bigr) \MW \CB^2\, \dTB + {} \notag \\
&&&\qquad \Bigl(
	    \SB (u_d - u_o) V_{12}^* V_{22}^* -
	    \CB (v_o - v_d) U_{12}^* U_{22}^*
	  \Bigr) \frac{\dMWsq}{\sqrt 2\,\MW} 
	\biggr\}\,, \notag \\
\Code{dMUE1} &\equiv \delta\mu &
	&= \frac 1{u_o v_o - u_d v_d}\biggl\{
	  U_{21}^* V_{21}^* \delta\mcha{1}^\OS -
	  U_{11}^* V_{11}^* \delta\mcha{2}^\OS - {} \\
&&&\qquad  \sqrt 2\, \Bigl(
	  \SB (u_d - u_o) V_{11}^* V_{21}^* +
	  \CB (v_o - v_d) U_{11}^* U_{21}^*
	  \Bigr) \MW \CB^2\, \dTB + {} \notag \\
&&&\qquad  \Bigl(
	  \CB (u_d - u_o) V_{11}^* V_{21}^* -
	  \SB (v_o - v_d) U_{11}^* U_{21}^*
	  \Bigr) \frac{\dMWsq}{\sqrt 2\,\MW} 
	\biggr\} \notag
\end{alignat}
\end{subequations}
where we use the short-hands
\begin{align*}
u_d &= U_{11}^* U_{22}^*\,, \quad u_o = U_{12}^* U_{21}^*\,, \quad
v_d  = V_{11}^* V_{22}^*\,, \quad v_o = V_{12}^* V_{21}^*\,, \\[1ex]
\delta N_n &= 2\CB^2\, \dTB \LP \SB N_{n3}^* + \CB N_{n4}^* \RP 
	\LP \MW N_{n2}^* - \MZ\SW N_{n1}^* \RP + {} \\
&\qquad	\LP \CB N_{n3}^* - \SB N_{n4}^* \RP
	\LP N_{n1}^* \LB \frac{\delta \MZ^2}{\MZ} \SW + 
	2\MZ\,\delta\SW\RB - N_{n2}^* \frac{\delta \MW^2}{\MW} \RP,
\notag \\
\delta\mneu{n}^\OS &= \ReTilde \LB {\mneu{n}}
	\Sigma_{\neu{}}^L(\mneu{n}^2) +
	\Sigma_{\neu{}}^{SL}(\mneu{n}^2) \RB_{nn}, \\
\delta\mcha{c}^\OS &= \ReTilde \LB 
	\frac{\mcha{c}}{2} \LP \Sigma_{\cha{}}^L(\mcha{c}^2) + 
	                       \Sigma_{\cha{}}^R(\mcha{c}^2) \RP +
	\Sigma_{\cha{}}^{SL}(\mcha{c}^2) \RB_{cc}.
\end{align*}


\subsubsection{CNN Schemes}

Renormalization schemes with one chargino $\chapm{c}$ and two 
neutralinos $\neu{n}$, $\neu{n'}$ taken as on-shell particles give 
better numerical stability in regions of $|\mu|\approx |M_2|$, as shown 
in \citere{onshellCNmasses}.

The on-shell conditions of the CNN schemes are analogous to those of the 
CCN schemes, see \citeres{Stop2decay,LHCxC,LHCxN}.  As above we impose 
conditions on the neutralino and chargino masses and solve for
$\delta M_1$, $\delta M_2$, and $\delta\mu$:
\begin{subequations}
\begin{alignat}{2}
\Code{dMino11} &\equiv \delta M_1 &
	&= \frac{A_{2,nn'} + 2 B_{2,nn'}\,\delta\mu}{C_{nn'}}\,, \\
\Code{dMino21} &\equiv \delta M_2 &
	&= -\frac{A_{1,nn'} + 2 B_{1,nn'}\,\delta\mu}{C_{nn'}}\,, \\
\Code{dMUE1} &\equiv \delta\mu &
	&= \frac 1{C_{nn'} U_{c2}^* V_{c2}^* -
	           2 B_{1,nn'} V_{c1}^* U_{c1}^*}
	  \biggl\{ U_{c1}^* V_{c1}^* A_{1,nn'} + {} \\
&&&\qquad\qquad	   C_{nn'} \Bigl[
	     \delta\mcha{c}^\OS -
	     \sqrt 2 (\CB\,U_{c1}^* V_{c2}^* - \SB\,V_{c1}^* U_{c2}^*)\,
	       \CB^2\,\dTB\,\MW - {} \notag \\
&&&\qquad\qquad\qquad\;
	     (\SB\,U_{c1}^* V_{c2}^* + \CB\,V_{c1}^* U_{c2}^*)
	       \frac{\delta\MW^2}{\sqrt{2}\MW} \Bigr]
	\biggr\}\,, \notag
\end{alignat}
\end{subequations}
with
\begin{align*}
A_{i,nn'} &= N_{n'i}^{*2} (\delta\mneu{n}^\OS + \delta N_n) -
	     N_{ni}^{*2} (\delta\mneu{n'}^\OS + \delta N_{n'})\,, \\
B_{i,nn'} &= N_{n'i}^{*2} N_{n3}^* N_{n4}^* -
	     N_{ni}^{*2} N_{n'3}^* N_{n'4}^*\,, \\
C_{nn'} &= N_{n1}^{*2} N_{n'2}^{*2} - N_{n'1}^{*2} N_{n2}^{*2}\,.
\end{align*}


\subsubsection{Discussion of Scheme Selection}
\label{sec:cnscheme}

In a recent analysis \cite{onshellCNmasses} it was emphasized that for a 
CCN scheme to yield numerically stable results it must be the most 
bino-like neutralino that is chosen on-shell.  \citere{Baro} further 
discusses the problem of large unphysical contributions due to a 
non-binolike lightest neutralino.

Note that the $Z$-factors also ensure that the external (stable) 
particle does not mix with other fields, which is one of the on-shell 
properties.  Which neutralino to take on-shell can be chosen (details 
below), currently the lightest one is the default.  For more discussion, 
see the Appendix of \citere{Stop2decay}.

A comparison of different renormalization schemes in the 
chargino/neutralino sector is given in \citere{LHCxN}.  The differences 
found with respect to another on-shell renormalization in the 
chargino/neutralino sector were small and of the expected size of 
two-loop contributions.

Special care has to be taken in the regions of the cMSSM parameter space 
where the gaugino--Higgsino mixing in the chargino sector is close to 
maximal, \ie where $|\mu|\approx |M_2|$.  Here $\delta M_2$ and 
$\delta\mu$ diverge as $(U^*_{11} U^*_{22} V^*_{11} V^*_{22} - U^*_{12} 
U^*_{21} V^*_{12} V^*_{21})^{-1}$ (cf.\ \refeqs{eq:dCha}) for the CCN 
schemes and the loop calculation does not yield a reliable result.  A 
similar singularity for $|\mu|\approx |M_2|$ arises also in the 
\Code{CNN[1,1,4]} and \Code{CNN[2,1,2]} schemes when $|M_1| < |M_2|$.  
These kind of divergences were also discussed in 
\citeres{dissAF,onshellCNmasses,bfmw}.

The choice of chargino/neutralino renormalization scheme is made through 
the variable \Code{\$InoScheme}, which must be set \emph{before} model 
initialization.  Allowed selections are \Code{CCN[$n$]} and 
\Code{CNN[$c$,$n$,$n'$]}; these may also be combined for run-time 
switching of the schemes, \eg
\begin{itemize}
\item \Code{\$InoScheme = CCN[1]}
--- fixed CCN scheme with on-shell $\neu{1}$.

This is currently the default but might change in the future based on 
experience with more calculations.

\item \Code{\$InoScheme = CNN[2,1,3]}
--- fixed CNN scheme with on-shell $\chapm{2}$, $\neu{1}$, $\neu{3}$.

\item \Code{\$InoScheme = IndexIf[\Var{cond}, CNN[2,1,3], CCN[1]]}

scheme chosen at run-time according to the logical Mathematica 
expression \Var{cond} which may contain model parameters, \eg
\Code{Abs[Abs[MUE]\,-\,Abs[Mino2]] < 50}. Note the use of \Code{IndexIf} 
instead of \Code{If}, necessary because \Code{If} has delayed evaluation 
of its arguments.

\item \Code{\$InoScheme = CCN[nbino]}
--- CCN scheme with the most bino-like neutralino on-shell, with
\Code{nbino} to be determined at run-time as the $n$ which maximizes 
$|N_{n1}|$.

Observe that a run-time switch of the renormalization scheme requires a 
corresponding transition of the affected parameters from one scheme to 
the other (not yet implemented) for a fully consistent interpretation of the
results.
\end{itemize}
Either scheme fixes three out of six chargino/neutralino masses to be 
on-shell.  The other three masses then acquire a finite shift.  It was 
shown in \citere{Stop2decay} that these shifts are small numerically.  
They are not implemented in the model file so far, which does not really 
count as an omission as propagators are parameterized with tree-level 
masses in canonical perturbation theory anyway.


\subsection{The Fermion Sector}
\label{sec:fermion}

We closely follow the renormalization of \citere{SbotRen,Stop2decay}, 
enlarged to the full fermion sector and extended to include external 
bottom quarks, too, for which the ``$\mb$, $\Ab$ \DRbar'' scheme 
proposed in \citere{SbotRen} is inappropriate. 

The fermion mass counterterms are defined as
\begin{equation}
\label{eq:dMf}
\Code{dMf1[\Vt,\,\Vg]} \equiv \delta\mf{tg}
	= U_{tg}\biggl\{\frac 12\ReTilde\Bigl[
	    \mf{tg}
	    \!\!\begin{aligned}[t]
	    &\mati{\SE{f_t}^L(\mf{tg}^2) +
	           \SE{f_t}^R(\mf{tg}^2)}_{gg} + \\
	    &\mati{\SE{f_t}^{SL}(\mf{tg}^2) +
	           \SE{f_t}^{SR}(\mf{tg}^2)}_{gg}
	\Bigr]\biggr\}\,.
	\end{aligned}
\end{equation}
Note that for the (massless) neutrinos ($t = 1$) this evaluates to
$\delta m_\nu = 0$.

The function $U_{tg}\equiv\Code{UVMf1[\Vt,\,\Vg]}$ allows to choose
between
\begin{subequations}
\label{eq:UVMf}
\begin{alignat}{2}
\Code{UVMf1[4,\,3]}~ &\Code{= UVDivergentPart} &\qquad
& \text{\DRbar\ renormalization for $\mb$ (default)}, \\
\Code{UVMf1[4,\,3]}~ &\Code{= Identity} &
& \text{on-shell (OS) renormalization for $\mb$}.
\end{alignat}
\end{subequations}
Unless the bottom quark appears in external states, the \DRbar\ 
prescription is preferred for $m_b$ \cite{SbotRen}.  The problems found 
in \citere{SbotRen} with an on-shell renormalization condition for $\mb$ 
(leading \eg to unphysically large contributions to $\delta A_b$) do not 
occur as long as there are no external sbottom quarks and the parameter 
$A_b$ is only needed at leading order.  An example for which both 
schemes would presumably fail is $b\bar b\to \Sbot_i\bar\Sbot_j$.  
More details on the definition of $\mb^\OS$ can be found in 
\citeres{SbotRen,Stop2decay}.

The fermion field renormalization constants are given by
\begin{subequations}
\label{eq:dZf}
\begin{alignat}{2}
\Code{dZfL1[\Vt,\,\Vg,\,\Vg]}
	&\equiv \mati{\dZm{f_t}^L + \dbZm{f_t}^L}_{gg} &
	&= \mati{\sigma_{gg}^L(f_t) + \tau_{gg}(f_t)}_{gg}\,, \\
\Code{dZbarfL1[\Vt,\,\Vg,\,\Vg]}
	&\equiv \mati{\dZm{f_t}^L + \dBZm{f_t}^L}_{gg} &
	&= \mati{\sigma_{gg}^L(f_t) - \tau_{gg}(f_t)}_{gg}\,, \\
\Code{dZfR1[\Vt,\,\Vg,\,\Vg]}
	&\equiv \mati{\dZm{f_t}^R + \dbZm{f_t}^R}_{gg} &
	&= \mati{\sigma_{gg}^R(f_t) - \tau_{gg}(f_t)}_{gg}\,, \\
\Code{dZbarfR1[\Vt,\,\Vg,\,\Vg]}
	&\equiv \mati{\dZm{f_t}^R + \dBZm{f_t}^R}_{gg} &
	&= \mati{\sigma_{gg}^R(f_t) + \tau_{gg}(f_t)}_{gg}\,,
\end{alignat}
\vspace*{-4ex}
\begin{align*}
\text{where}\quad
\sigma_{ii}^X(f) &= -\ReDiag\Bigl[
	\SE{f_i}^X(\mf{i}^2) +
	\!\begin{aligned}[t]
	&\mf{i}^2 \LP\SE{f}^{\prime L}(\mf{i}^2) +
	             \SE{f}^{\prime R}(\mf{i}^2)\RP + {} \\
	&\mf{i} \LP\SE{f}^{\prime SL}(\mf{i}^2) +
	           \SE{f}^{\prime SR}(\mf{i}^2)\RP
	\Big]\,,
	\end{aligned} \\
\tau_{ii}(f) &= \begin{cases}
  \dfrac 1{2\mf{i}}\LP
	\ReDiag \LB\SE{f}^{SL}(\mf{i}^2) -
	           \SE{f}^{SR}(\mf{i}^2)\RB\RP &
  \text{if $\mf{i}\neq 0$}\,, \\
  0 & \text{if $\mf{i} = 0$}\,.
\end{cases}
\end{align*}
As we presently do not take neutrino masses into account, the following
off-diagonal entries ($g\neq g'$) appear only in the quark sector ($t = 
3, 4$) and only for non-trivial CKM matrix:
\begin{alignat}{2}
\Code{dZfL1[\Vt,\,\Vg,\,\Vgp]}
	&\equiv \mati{\dZm{f_t}^L + \dbZm{f_t}^L}_{gg'} &
	&= \mati{\sigma_{gg'}^{L,R,SR,SL}(f_t)}_{gg'}\,, \\
\Code{dZbarfL1[\Vt,\,\Vg,\,\Vgp]}
	&\equiv \mati{\dZm{f_t}^L + \dBZm{f_t}^L}_{gg'} &
	&= \mati{\sigma_{g'g}^{L,R,SL,SR}(f_t)}_{gg'}\,, \\
\Code{dZfR1[\Vt,\,\Vg,\,\Vgp]}
	&\equiv \mati{\dZm{f_t}^R + \dbZm{f_t}^R}_{gg'} &
	&= \mati{\sigma_{gg'}^{R,L,SL,SR}(f_t)}_{gg'}\,, \\
\Code{dZbarfR1[\Vt,\,\Vg,\,\Vgp]}
	&\equiv \mati{\dZm{f_t}^R + \dBZm{f_t}^R}_{gg'} &
	&= \mati{\sigma_{g'g}^{R,L,SR,SL}(f_t)}_{gg'}\,,
\end{alignat}
\vspace*{-4ex}
\begin{align*}
\text{where}\quad
\sigma_{ij}^{X,Y,SX,SY}(f)
	&= \frac 2{\mf{i}^2 - \mf{j}^2} \ReOffDiag\Bigl[
	\mf{j}\bigl(
	\!\begin{aligned}[t]
	&\mf{j}\SE{f}^X(\mf{j}^2) +
	 \mf{i}\SE{f}^Y(\mf{j}^2)\bigr) + {} \\
	&\mf{j}\SE{f}^{SX}(\mf{j}^2) +
	 \mf{i}\SE{f}^{SY}(\mf{j}^2)\Bigr]\,.
	\end{aligned}
\end{align*}
\end{subequations}

The CKM matrix $V_{ij}$ receives counterterms \cite{denner}
\begin{alignat}{2}
\label{eq:dCKM}
\Code{dCKM1[\Vi,\,\Vj]} &\equiv \delta V_{ij} &
	&= \frac 14\sum_{g = 1}^3\LP
	   \mati{\dZm{u}^L - \dZm{u}^{L\dagger}}_{ig} V_{gj} -
	   V_{ig}\mati{\dZm{d}^L - \dZm{d}^{L\dagger}}_{gj}\RP.
\end{alignat}

A note on CKM mixing: The model file is presently limited to minimal 
flavor violation in the sfermion sector (see \refse{sec:sfermion}), 
which means that for non-trivial CKM matrix there is a slight imbalance 
between fermions and sfermions; for example, the $b$-quark has an 
admixture from $d$ and $s$ while the $\tilde b$ does not.  Because this 
violates delicate supersymmetric relations, processes involving squarks 
(in particular external ones) may not become finite and we have 
therefore chosen to make \Code{\$CKM = False} ($V_{ij} = \delta_{ij}$, 
$\delta V_{ij} = 0$) the default.  CKM mixing can be turned on with 
\Code{\$CKM = True} (set \emph{before} model initialization).


\subsection{The Sfermion Sector}
\label{sec:sfermion}

In the absence of non-minimal flavor violation, the sfermion mass matrix
is given by \cite{HaK85,GuH86}
\begin{equation}
\matr{M}^2_{\Sf_{tg}} = \begin{pmatrix}
  \bigl(\matr{M}^2_{L,f_t}\bigr)_{gg} + \mf{tg}^2
	& \mf{tg} \bigl(\matr{X}_{f_t}\bigr)_{gg}^* \\
  \mf{tg} \bigl(\matr{X}_{f_t}\bigr)_{gg}
	& \bigl(\matr{M}^2_{R,f_t}\bigr)_{gg} + \mf{tg}^2
\end{pmatrix}
\end{equation}
where
\begin{align*}
\matr{M}^2_{L,f_t} &= \MZ^2 (I_3^{f_t} - Q_{f_t}\SW^2) \CBB + 
\begin{cases}
\matr{M}^2_{\tilde L}
	& \text{for left-handed sleptons ($t = 1, 2$)}\,, \\
\matr{M}^2_{\tilde Q}
	& \text{for left-handed squarks ($t = 3, 4$)}\,,
\end{cases} \\
\matr{M}^2_{R,f_t} &= \MZ^2 Q_{f_t}\SW^2 \CBB + 
\begin{cases}
\matr{M}^2_{\tilde E}
	& \text{for right-handed sleptons ($t = 2$)}\,, \\
\matr{M}^2_{\tilde U}
	& \text{for right-handed $u$-type squarks ($t = 3$)}\,, \\
\matr{M}^2_{\tilde D}
	& \text{for right-handed $d$-type squarks ($t = 4$)}\,, \\
\end{cases} \\
\matr{X}_{f_t} &= \matr{A}_{f_t} - \mu^*
\begin{cases}
1/\TB & \text{for isospin-up sfermions ($t = 3$)}\,, \\
\TB & \text{for isospin-down sfermions ($t = 2, 4$)}\,.
\end{cases}
\end{align*}
The soft-SUSY-breaking parameters $\matr{M}^2_{\tilde L,\tilde Q,
\tilde E,\tilde U,\tilde D}$ and $\matr{A}_f$ are $3\times 3$ matrices 
in flavor space whose off-diagonal entries are zero in the minimally
flavor-violating MSSM.

The mass matrix is diagonalized by a unitary transformation
${\matr{U}}_\Sf$,
\begin{align}
\matr{U}_\Sf\matr{M}^2_\Sf\matr{U}_\Sf^\dagger =
\begin{pmatrix}
  \msf1^2 & 0 \\
  0 & \msf2^2
\end{pmatrix}, \qquad
\matr{U}_\Sf = \begin{pmatrix}
  U^\Sf_{11} & U^\Sf_{12} \\
  U^\Sf_{21} & U^\Sf_{22}
\end{pmatrix}.
\end{align}


\subsubsection{The Squark Sector}
\label{sec:squark}

We renormalize the up-type squarks ($\Su = \{\tilde u, \tilde c, \tilde 
t\}$) on-shell (OS).  For the down-type squarks ($\Sd = \{\tilde d, 
\tilde s, \tilde b\}$) we follow the discussion in Sect.~4 of
\citere{SbotRen}.

Also the down-type squark masses could in principle both be renormalized 
on-shell (option $\mathcal{O}2$ of \citere{SbotRen}).  They would then 
have to be computed from a mass matrix with shifted $M_{\tilde Q}^2$, 
however, which is not entirely straightforward to implement in practice.  
We therefore set only one down-type squark mass $\smash{(\msd{ig})}$ 
on-shell and continue to work with a tree-level mass matrix.  The 
one-loop-corrected on-shell value of the remaining mass 
$\smash{\msd{jg}}$ then acquires a shift involving the dependent mass 
counterterm $\delta\msd{jg}^2$,
\begin{equation}
\bigl(\msd{jg}^\OS\bigr)^2
= \msd{jg}^2 + \delta\msd{jg}^2 -
  \ReTilde\mati{\SE{\Sd_g}(\msd{jg}^2)}_{jj}\,.
\end{equation}

{\samepage
For the other parameters two scheme choices are available:
\begin{itemize}
\item
By default we apply the ``$m_b, A_b$ \DRbar'' scheme of 
\citeres{SbotRen,Stop2decay}.  (For completeness we note that a slight 
extension of this scheme was used there.)  We shall refer to this as the 
\textbf{mixed scheme} -- the original term ``$m_b, A_b$ \DRbar'' is not 
quite accurate as we allow the bottom quark to be chosen on-shell, 
see \refeq{eq:UVMf}.

\item
Alternately, an \textbf{on-shell scheme} can be used for the squark 
sector.  Despite the name also here only one down-type squark mass is 
on-shell but the off-diagonal `mixing-angle' counterterm $\delta 
Y_{\Fd_g}$ is fixed by an on-shell-type condition and the trilinear 
coupling's counterterm $\delta A_{\Fd_g}$ becomes a dependent quantity.
\end{itemize}}

The scheme affecting sfermions $\Sd_g$ is chosen with the variable 
\Code{\$SfScheme[4,\,\Vg]}:
\begin{subequations}
\label{eq:sqscheme}
\begin{alignat}{2}
\Code{\$SfScheme[4,\,\Vg]}~ &\Code{= DR[1]} &\qquad
	& \text{mixed scheme with $\msd{1g}$ OS, $A_{\Fd_g}$ \DRbar}, \\
\Code{\$SfScheme[4,\,\Vg]}~ &\Code{= DR[2]} &\qquad
	& \text{mixed scheme with $\msd{2g}$ OS, $A_{\Fd_g}$ \DRbar\ (default)}, \\[1ex]
\Code{\$SfScheme[4,\,\Vg]}~ &\Code{= OS[1]} &\qquad
	& \text{on-shell scheme with $\msd{1g}$ OS, $Y_{\Fd_g}$ OS}, \\
\Code{\$SfScheme[4,\,\Vg]}~ &\Code{= OS[2]} &\qquad
	& \text{on-shell scheme with $\msd{2g}$ OS, $Y_{\Fd_g}$ OS}.
\end{alignat}
\end{subequations}
As for \Code{\$InoScheme}, conditional scheme selection (\eg to avoid 
regions of numerical instability) can be set up as in
\Code{IndexIf[\Var{cond},\,DR[1],\,DR[2]]}.

In the following, \Vi\ denotes the index of \Code{OS} and \Vj\ the other
($j = 3 - i$).  The index \Vs\ runs over both values 1, 2.  The two
up-type and the chosen down-type sfermion are on-shell,
\begin{subequations}
\label{eq:dMSq}
\begin{alignat}{2}
\Code{dMSfsq1[\Vs,\,\Vs,\,3,\,\Vg]} &\equiv {} & \delta\msu{sg}^2
	&= \ReTilde\mati{\SE{\Su_g}(\msu{sg}^2)}_{ss}\,, \\
\Code{dMSfsq1[\Vi,\,\Vi,\,4,\,\Vg]} &\equiv {} & \delta\msd{ig}^2
	&= \ReTilde\mati{\SE{\Sd_g}(\msd{ig}^2)}_{ii}\,.
\end{alignat}
The up-type off-diagonal mass-matrix entries receive counterterms 
\cite{mhcMSSM2L,dissHR,SbotRen}
\begin{alignat}{2}
\Code{dMSfsq1[1,\,2,\,3,\,\Vg]}
	&\equiv {} & \delta Y_{\Fu_g}
	&= \frac 12\ReTilde\mati{\SE{\Su_g}(\msu{1g}^2) +
	                         \SE{\Su_g}(\msu{2g}^2)}_{12}\,, \\
\Code{dMSfsq1[2,\,1,\,3,\,\Vg]}
	&= {} & \delta Y_{\Fu_g}^*
	&= \frac 12\ReTilde\mati{\SE{\Su_g}(\msu{1g}^2) +
	                         \SE{\Su_g}(\msu{2g}^2)}_{21}\,.
\end{alignat}
\end{subequations}
For clarity of notation we furthermore define the auxiliary constants
\begin{subequations}
\begin{align}
\Code{dMsq11Sf1[4,\,\Vg]} \equiv \delta M_{\Sd_g,11}^2
	&= |U^{\Su_g}_{11}|^2 \delta\msu{1g}^2 +
	   |U^{\Su_g}_{12}|^2 \delta\msu{2g}^2 -
	   2 \Re\bigl[U^{\Su_g}_{22} U^{\Su_g*}_{12}
	     \delta Y_{\Fu_g}\bigr] - {} \\
&\qquad 2\mfu{g}\delta\mfu{g} + 2\mfd{g}\delta\mfd{g} -
	  \CBB\,\delta\MW^2 + 4\MW^2\CB^3\SB\,\dTB\,, \notag \\
\Code{dMsq12Sf1[4,\,\Vg]} \equiv \delta M_{\Sd_g,12}^2
	&= \mfd{g} (\delta A_{\Fd_g}^* - \mu\,\dTB - \TB\,\delta\mu) +
	   (A_{\Fd_g}^* - \mu\,\TB)\,\delta\mfd{g}\,.
\end{align}
\end{subequations}
For the bottom quark two options are possible: $\delta\mb = 
\delta\mb^{\DRbar}$ or $\delta\mb = \delta\mb^\OS$, see 
\refse{sec:fermion}.

In the \textbf{mixed scheme} the dependent mass counterterm is
\begin{subequations}
\label{eq:dMSd}
\begin{align}
\label{eq:dMSd11}
\Code{dMSfsq1[\Vj,\,\Vj,\,4,\,\Vg]} &\equiv \delta\msd{jg}^2
	= \frac 1{|U^{\Sd_g}_{1j}|^2}
	\Bigl\{
	  |U^{\Sd_g}_{1i}|^2 \delta\msd{ig}^2 + {} \\
&\kern 2em
	  (i - j) \bigl( 2\Re\bigl[ U^{\Sd_g}_{11} U^{\Sd_g*}_{12}
	    \delta M_{\Sd_g,12}^2\bigr] +
	  \bigl(|U^{\Sd_g}_{11}|^2 - |U^{\Sd_g}_{12}|^2\bigr)
	    \rlap{$\delta M_{\Sd_g,11}^2 \bigr) \Bigr\}\,,$} \notag
\intertext{and the down-type off-diagonal mass counterterms are related as}
\Code{dMSfsq1[1,\,2,\,4,\,\Vg]} &\equiv \delta Y_{\Fd_g}
	= \frac 1{|U^{\Sd_g}_{11}|^2 - |U^{\Sd_g}_{12}|^2}
	\Bigl\{
	  U^{\Sd_g}_{11} U^{\Sd_g*}_{21}
	  \bigl(\delta\msd{1g}^2 - \delta\msd{2g}^2\bigr) + {} \\[-1ex]
&\kern 12em
	  U^{\Sd_g}_{11} U^{\Sd_g*}_{22} \delta M_{\Sd_g,12}^2 -
	  U^{\Sd_g}_{12} U^{\Sd_g*}_{21} \delta M_{\Sd_g,12}^{2*}
	\Bigr\}\,, \notag \\
\Code{dMSfsq1[2,\,1,\,4,\,\Vg]} &= \delta Y_{\Fd_g}^*\,.
\intertext{In the \textbf{on-shell scheme} we have instead}
\Code{dMSfsq1[\Vj,\,\Vj,\,4,\,\Vg]} &\equiv \delta\msd{jg}^2
	= \frac 1{|U^{\Sd_g}_{1j}|^2} \Bigl\{
	\!\begin{aligned}[t]
	& {-}|U^{\Sd_g}_{1i}|^2 \delta\msd{ig}^2 +
	  2\Re\bigl[U^{\Sd_g}_{22} U^{\Sd_g*}_{12}
	    \delta Y_{\Fd_g}\bigr] + \\
	& \delta M_{\Sd_g,11}^2	\Bigr\}\,,
	\end{aligned} \\
\Code{dMSfsq1[1,\,2,\,4,\,\Vg]} &\equiv \delta Y_{\Fd_g}
	= \frac 12\ReTilde\mati{\SE{\Sd_g}(\msd{1g}^2) +
	                        \SE{\Sd_g}(\msd{2g}^2)}_{12}\,, \\
\Code{dMSfsq1[2,\,1,\,4,\,\Vg]} &= \delta Y_{\Fd_g}^*
	= \frac 12\ReTilde\mati{\SE{\Sd_g}(\msd{1g}^2) +
	                        \SE{\Sd_g}(\msd{2g}^2)}_{21}\,.
\end{align}
\end{subequations}
In both schemes the trilinear couplings
$A_{q_{tg}}\equiv\mati{\matr{A}_{q_t}}_{gg}$ are renormalized by
\begin{subequations}
\label{eq:dAq}
\begin{alignat}{2}
\Code{dAf1[3,\,\Vg,\,\Vg]} &\equiv \delta A_{\Fu_g} &
	&= \frac 1{\mfu{g}} \Bigl[
	\rlap{$\begin{aligned}[t]
	& U^{\Su_g}_{11} U^{\Su_g*}_{12}
	    (\delta\msu{1g}^2 - \delta\msu{2g}^2) + {} \\
	& U^{\Su_g}_{11} U^{\Su_g*}_{22} \delta Y_{\Fu_g}^* +
	  U^{\Su_g*}_{12} U^{\Su_g}_{21} \delta Y_{\Fu_g} -
	  \bigl(A_{\Fu_g} - \mu^*/\TB\bigr)\,\delta\mfu{g}
	\rlap{$\Bigr] + {}$}
	\end{aligned}$} \\
&&	&\qquad \delta\mu^*/\TB - \mu^* \dTB/\TB^2\,, \notag \\[1ex]
\Code{dAf1[4,\,\Vg,\,\Vg]} &\equiv \delta A_{\Fd_g} &
	&= \biggl\{\frac 1{\mfd{g}} \Bigl[
	\begin{aligned}[t]
	& U^{\Sd_g}_{11} U^{\Sd_g*}_{12}
	    (\delta\msd{1g}^2 - \delta\msd{2g}^2) + {} \\
	& U^{\Sd_g}_{11} U^{\Sd_g*}_{22} \delta Y_{\Fd_g}^* +
	  U^{\Sd_g*}_{12} U^{\Sd_g}_{21} \delta Y_{\Fd_g} -
	  \bigl(A_{\Fd_g} - \mu^*\TB\bigr)\,\delta\mfd{g}
	\rlap{$\Bigr] + {}$}
	\end{aligned} \\
&&	&\qquad\, \TB\,\delta\mu^* + \mu^*\dTB
	\biggr\}_{\mathrm{[div]}}\,, \notag
\end{alignat}
\end{subequations}
where the subscripted [div] means to take the divergent part in the 
mixed scheme only, to effect \DRbar\ renormalization of $A_{\Fd_g}$ 
\cite{SbotRen}.

The sfermion $Z$-factors are derived in the OS scheme.  The OS scheme 
has nothing to say on the imaginary parts of the diagonal $Z$-factors 
and since they contain no divergences, we implicitly set them to zero 
below.
\begin{subequations}
\label{eq:dZSq}
\begin{alignat}{2}
\Code{dZSf1[\Vs,\,\Vs,\,3,\,\Vg]}
	&\equiv \mati{\dZm{\Su_g} + \dbZm{\Su_g}}_{ss} &
	&= -\ReDiag\mati{\Sigma'_{\Su_g}(\msu{sg}^2)}_{ss}\,, \\
\Code{dZbarSf1[\Vs,\,\Vs,\,3,\,\Vg]}
	&= \rlap{\Code{dZSf1[\Vs,\,\Vs,\,3,\,\Vg]}\,,} \\[1ex]
\Code{dZSf1[\Vs,\,\Vs,\,4,\,\Vg]}
	&\equiv \mati{\dZm{\Sd_g} + \dbZm{\Sd_g}}_{ss} &
	&= -\ReDiag\mati{\Sigma'_{\Sd_g}(\msd{sg}^2)}_{ss}\,, \\
\Code{dZbarSf1[\Vs,\,\Vs,\,4,\,\Vg]}
	&= \rlap{\Code{dZSf1[\Vs,\,\Vs,\,4,\,\Vg]}\,,} \\[2ex]
\Code{dZSf1[1,\,2,\,3,\,\Vg]}
	&\equiv \mati{\dZm{\Su_g} + \dbZm{\Su_g}}_{12} &
	&= +\sigma^{12}_2(\Su_g)\,, \\
\Code{dZbarSf1[1,\,2,\,3,\,\Vg]}
	&\equiv \mati{\dZm{\Su_g} + \dBZm{\Su_g}}_{12} &
	&= -\sigma^{21}_2(\Su_g)\,, \\
\Code{dZSf1[2,\,1,\,3,\,\Vg]}
	&\equiv \mati{\dZm{\Su_g} + \dbZm{\Su_g}}_{21} &
	&= +\sigma^{21}_1(\Su_g)\,, \\
\Code{dZbarSf1[2,\,1,\,3,\,\Vg]}
	&\equiv \mati{\dZm{\Su_g} + \dBZm{\Su_g}}_{21} &
	&= -\sigma^{12}_1(\Su_g)\,, \\[2ex]
\Code{dZSf1[1,\,2,\,4,\,\Vg]}
	&\equiv \mati{\dZm{\Sd_g} + \dbZm{\Sd_g}}_{12} &
	&= +\sigma^{12}_2(\Sd_g)\,, \\
\Code{dZbarSf1[1,\,2,\,4,\,\Vg]}
	&\equiv \mati{\dZm{\Sd_g} + \dBZm{\Sd_g}}_{12} &
	&= -\sigma^{21}_2(\Sd_g)\,, \\
\Code{dZSf1[2,\,1,\,4,\,\Vg]}
	&\equiv \mati{\dZm{\Sd_g} + \dbZm{\Sd_g}}_{21} &
	&= +\sigma^{21}_1(\Sd_g)\,, \\
\Code{dZbarSf1[2,\,1,\,4,\,\Vg]}
	&\equiv \mati{\dZm{\Sd_g} + \dBZm{\Sd_g}}_{21} &
	&= -\sigma^{12}_1(\Sd_g)\,,
\end{alignat}
\vspace*{-3ex}
\begin{align*}
\text{where}\quad
\sigma^{12}_i(\tilde f) &= \frac 2{\msf1^2 - \msf2^2}
  \bigl(\ReOffDiag\mati{\SE{\Sf}(\msf{i}^2)}_{12} - \delta Y_f\bigr)\,, \\
\sigma^{21}_i(\tilde f) &= \frac 2{\msf2^2 - \msf1^2}
  \bigl(\ReOffDiag\mati{\SE{\Sf}(\msf{i}^2)}_{21} - \delta Y_f^*\bigr)\,.
\end{align*}
\end{subequations}
CPT invariance requires that $\ReTilde\mati{\SE{\Sf}(p^2)}_{ss'} =
\bigl(\ReTilde\mati{\SE{\Sf}(p^2)}_{s's}\bigr)^*$ and it is indeed true 
that, leaving out the absorptive parts, $\mati{\dZm{\Sf}}_{ss'} = 
\mati{\dZm{\Sf}}_{s's}^*$.


\subsubsection{The Slepton Sector}
\label{sec:slepton}

The discussion of the slepton renormalization can be fairly brief as we 
merely adapt the squark-sector results 
\cite{SbotRen,Stop2decay,LHCxC,LHCxN} to account for the absence of 
neutrino masses (and hence only a single sneutrino).  We restrict 
ourselves to the on-shell scheme which we found to give numerically
stable results up to relatively large values of $\TB$.
\begin{subequations}
\label{eq:slscheme}
\begin{alignat}{2}
\Code{\$SfScheme[2,\,\Vg]}~ &\Code{= OS[1]} &\qquad
	& \text{on-shell scheme with $\mse{1g}$ OS, $Y_{\Fe_g}$ OS}, \\
\Code{\$SfScheme[2,\,\Vg]}~ &\Code{= OS[2]} &\qquad
	& \text{on-shell scheme with $\mse{2g}$ OS, $Y_{\Fe_g}$ OS}.
\end{alignat}
\end{subequations}
As before, $i$ denotes the index of \Code{OS} and $j = 3 - i$ the
other.

We fix two out of the three slepton masses on-shell ($\Se = \{\tilde e, 
\tilde\mu, \tilde\tau\}$) \cite{hr,Stau2decay},
\begin{subequations}
\label{eq:dMSl}
\begin{alignat}{2}
\Code{dMSfsq1[1,\,1,\,1,\,\Vg]} &\equiv \delta\msnu{1g}^2 &
	&= \ReTilde\mati{\SE{\Snu_g}(\msnu{1g}^2)}_{11}\,, \\
\Code{dMSfsq1[\Vi,\,\Vi,\,2,\,\Vg]} &\equiv \delta\mse{ig}^2 &
	&= \ReTilde\mati{\SE{\Se_g}(\mse{ig}^2)}_{ii}\,.
\end{alignat}
The non-diagonal entries of the selectron-type mass matrix are 
determined by \cite{mhiggsFDalbals,hr,SbotRen}
\begin{align}
\Code{dMSfsq1[1,\,2,\,2,\,\Vg]} &\equiv \delta Y_{\Fe_g}
	= \frac 12\ReTilde\mati{\SE{\Se_g}(\mse{1g}^2) +
	                        \SE{\Se_g}(\mse{2g}^2)}_{12}\,, \\
\Code{dMSfsq1[2,\,1,\,2,\,\Vg]} &\equiv \delta Y_{\Fe_g}^*
	= \frac 12\ReTilde\mati{\SE{\Se_g}(\mse{1g}^2) +
	                        \SE{\Se_g}(\mse{2g}^2)}_{21}\,,
\intertext{and the counterterm for the remaining dependent mass
$\mse{jg}^2$ is}
\Code{dMSfsq1[\Vj,\,\Vj,\,2,\,\Vg]} &\equiv \delta\mse{jg}^2
	= \frac 1{|U^{\Se_g}_{1j}|^2} \Bigl\{
	\begin{aligned}[t]
	& {-}|U^{\Se_g}_{1i}|^2 \delta\mse{ig}^2 +
	  2 \Re\big[U^{\Se_g*}_{12} U^{\Se_g}_{22}
	    \delta Y_{\Fe_g}\bigr] + {} \\
	& \delta M_{\Se_g,11}^2 \Bigr\}
	\end{aligned}
\end{align}
\end{subequations}
where we used the analogous auxiliary quantity
\begin{subequations}
\begin{align}
\Code{dMsq11Sf1[2,\,\Vg]} \equiv \delta M_{\Se_g,11}^2
	&= \delta\msnu{1g}^2 + 2\mfe{g}\delta\mfe{g} -
	   \CBB\delta\MW^2 + 4\MW^2\CB^3\SB\dTB\,.
\end{align}
\end{subequations}

The trilinear couplings are renormalized by 
\begin{alignat}{2}
\label{eq:dAe}
\Code{dAf1[2,\,\Vg,\,\Vg]} &\equiv \delta A_{\Fe_g} &
	&= \frac 1{\mfe{g}}\bigl[
	\begin{aligned}[t]
	& U^{\Se_g}_{11} U^{\Se_g*}_{12}
	  (\delta\mse{1g}^2 - \delta\mse{2g}^2) + {} \\
	& U^{\Se_g}_{11} U^{\Se_g*}_{22} \delta Y_{\Fe_g}^* +
	  U^{\Se_g*}_{12} U^{\Se_g}_{21} \delta Y_{\Fe_g} -
	  (A_{\Fe_g} - \mu^*\TB)\,\delta\mfe{g}
	\rlap{$\bigr] + {}$}
	\end{aligned} \\
&&	&\qquad \delta\mu^*\TB + \mu^*\dTB\,. \notag
\end{alignat}

The slepton $Z$-factors are:
\begin{subequations}
\label{eq:dZSl}
\begin{alignat}{2}
\Code{dZSf1[1,\,1,\,1,\,\Vg]}
	&\equiv \mati{\dZm{\Snu_g} + \dbZm{\Snu_g}}_{11} &
	&= -\ReDiag\mati{\Sigma'_{\Snu_g}(\msnu{1g}^2)}_{11}\,, \\
\Code{dZbarSf1[1,\,1,\,1,\,\Vg]}
	&= \rlap{\Code{dZSf1[1,\,1,\,1,\,\Vg]}\,,} \\[1ex]
\Code{dZSf1[\Vs,\,\Vs,\,2,\,\Vg]}
	&\equiv \mati{\dZm{\Se_g} + \dbZm{\Se_g}}_{ss} &
	&= -\ReDiag\mati{\Sigma'_{\Se_g}(\mse{sg}^2)}_{ss}\,, \\
\Code{dZbarSf1[\Vs,\,\Vs,\,2,\,\Vg]}
	&= \rlap{\Code{dZSf1[\Vs,\,\Vs,\,2,\,\Vg]}\,,} \\[2ex]
\Code{dZSf1[1,\,2,\,2,\,\Vg]}
	&\equiv \mati{\dZm{\Se_g} + \dbZm{\Se_g}}_{12} &
	&= +\sigma^{12}_2(\Se_g)\,, \\
\Code{dZbarSf1[1,\,2,\,2,\,\Vg]}
	&\equiv \mati{\dZm{\Se_g} + \dBZm{\Se_g}}_{12} &
	&= -\sigma^{21}_2(\Se_g)\,, \\
\Code{dZSf1[2,\,1,\,2,\,\Vg]}
	&\equiv \mati{\dZm{\Se_g} + \dbZm{\Se_g}}_{21} &
	&= +\sigma^{21}_1(\Se_g)\,, \\
\Code{dZbarSf1[2,\,1,\,2,\,\Vg]}
	&\equiv \mati{\dZm{\Se_g} + \dBZm{\Se_g}}_{21} &
	&= -\sigma^{12}_1(\Se_g)\,,
\end{alignat}
\vspace*{-4ex}
\begin{align*}
\text{where}\quad
\sigma^{12}_i(\tilde f) &= \frac 2{\msf1^2 - \msf2^2}
  \bigl(\ReOffDiag\mati{\SE{\Sf}(\msf{i}^2)}_{12} - \delta Y_f\bigr)\,, \\
\sigma^{21}_i(\tilde f) &= \frac 2{\msf2^2 - \msf1^2}
  \bigl(\ReOffDiag\mati{\SE{\Sf}(\msf{i}^2)}_{21} - \delta Y_f^*\bigr)\,.
\end{align*}
\end{subequations}


\subsection{The Gluino Sector}
\label{sec:gluino}

The gluino sector is determined by the soft-breaking gluino mass 
parameter $M_3 = |M_3|\re^{\ri\phigl}$.  A real gluino mass $\mgl = 
|M_3|$ is obtained through the transformation
\begin{align}
\omega_\pm\gl^a \to \re^{\mp\ri\phigl/2} \omega_\pm\gl^a
\end{align}
of the gluino field $\gl^a$ which moves the gluino phase into the 
couplings, where it appears as $\Code{SqrtEGl}\equiv\re^{\ri\phigl/2}$.

The gluino is renormalized on-shell \cite{Stop2decay,Gluinodecay},
\begin{equation}
\label{eq:dMGl}
\Code{dMGl1} \equiv \delta M_3
	= \frac 12\ReTilde\Bigl[
	   \mgl\bigl(\SE{\gl}^L(\mgl^2) + \SE{\gl}^R(\mgl^2)\bigr) +
	   \SE{\gl}^{SL}(\mgl^2) +\SE{\gl}^{SR}(\mgl^2)
	\Bigr]\re^{\ri\phigl}\,,
\end{equation}
\vspace*{-2.5ex}
\begin{subequations}
\label{eq:dZGl}
\begin{alignat}{2}
\Code{dZGlL1} &\equiv \dZ{\gl}^L + \dbZ{\gl}^L &
	&= \sigma^L(\gl) + \tau(\gl) - \ri\,\delta\phigl\,, \\
\Code{dZbarGlL1}
	&= \rlap{\Code{dZGlR1}\,,} \\
\Code{dZGlR1} &\equiv \dZ{\gl}^R + \dbZ{\gl}^R &
	&= \sigma^R(\gl) - \tau(\gl) - \ri\,\delta\phigl\,, \\
\Code{dZbarGlR1}
	&= \rlap{\Code{dZGlL1}\,,}
\end{alignat}
\vspace*{-3ex}
\begin{align*}
\text{where}\quad
\sigma^X(\gl) &= -\ReDiag\Bigl[
	\SE{\gl}^X(\mgl^2) +
	\!\begin{aligned}[t]
	&\mgl^2 \LP\SE{\gl}^{\prime L}(\mgl^2) +
	           \SE{\gl}^{\prime R}(\mgl^2)\RP + {} \\
	&\mgl \LP\SE{\gl}^{\prime SL}(\mgl^2) +
	         \SE{\gl}^{\prime SR}(\mgl^2)\RP
	\Big]\,,
	\end{aligned} \\
\tau(\gl) &= \frac 1{2\mgl}\LP
	\ReDiag \LB\SE{\gl}^{SL}(\mgl^2) -
	           \SE{\gl}^{SR}(\mgl^2)\RB\RP.
\end{align*}
\end{subequations}
We choose $\delta\phigl = 0$ with a rationale as in the quark case with 
no flavor violation: There, the Yukawa coupling can be made real by a 
redefinition of the quark fields and a complex $Z$-factor keeps it that 
way also at one-loop order.  For the gluinos the phase still appears in 
the Lagrangian after field redefinition but this phase factor can be 
considered a `transformation matrix' and does not obtain a counterterm.



\subsection{The Gluon Sector}
\label{sec:gluon}

We use \DRbar\ renormalization for $\alpha_s$ and the gluon field,
\begin{alignat}{2}
\label{eq:dZgs}
\Code{dZgs1} &\equiv \dZ{\alpha_s} &
	&= \frac 12\dZ{gg}\,, \\
\label{eq:dZGG}
\Code{dZGG1} &\equiv \dZ{gg} &
	&= -\ReTilde\Sigma^{\prime T}_G(0)\ddiv\,.
\end{alignat}


\section{Usage}
\label{sec:usage}

The FeynArts--FormCalc system works in three stages, as sketched in the 
following diagram.  A more detailed discussion of the interplay between 
FeynArts and FormCalc can be found in \citere{acat10}.
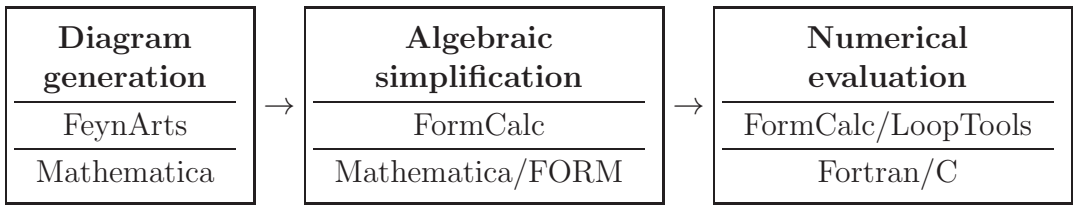
\begin{equation*}
\fbox{%
\begin{tabular}{c}
\textbf{Diagram} \\[-.5ex]
\textbf{generation} \\ \hline
FeynArts \\ \hline
Mathematica
\end{tabular}
}\to\fbox{%
\begin{tabular}{c}
\textbf{Algebraic} \\[-.5ex]
\textbf{simplification} \\ \hline
FormCalc \\ \hline
Mathematica/FORM
\end{tabular}
}\to\fbox{%
\begin{tabular}{c}
\textbf{Numerical} \\[-.5ex]
\textbf{evaluation} \\ \hline
FormCalc/LoopTools \\ \hline
Fortran/C
\end{tabular}
}
\end{equation*}
The model file can of course be used even if one does not wish to 
continue the calculation with FormCalc.

\subsection{Diagram Generation}

The \Code{MSSMCT.mod} and \Code{SQCDCT.mod} model files are used like 
any other FeynArts model files.  At present, they \emph{cannot} be used 
together with the FeynArts add-on model file \Code{FV.mod} 
\cite{interplay}, which extends the minimal flavor mixing to full 
$6\times 6$ sfermion mixing.  Extension of the sfermion renormalization 
of \refse{sec:sfermion} to the non-minimal case is work in progress, 
however.

The model file can be influenced by redefining the following symbols:
\begin{itemize}
\item
Wave-function correction factors $\dbZ{}$ are omitted by setting
\Code{ReDiag} (for diagonal $\dbZ{ii}$) and/or \Code{ReOffDiag} (for 
off-diagonal $\dbZ{ij}$) to \Code{ReTilde}, see \refeqs{eq:rediag}.

\item
An on-shell bottom mass is chosen by setting \Code{UVMf1[4,\,3]} to 
\Code{Identity}, see \refeqs{eq:UVMf}.

\item
The variable \Code{\$MHpInput} (set \emph{before} model initialization) 
decides whether $\MA$ (\Code{False}) or $\MHp$ (\Code{True}, default) 
are taken as input, see \refeqs{eq:MHpInput}.  Regarding the choice of 
tree-level vs.\ loop-corrected Higgs masses as discussed below 
\refeq{eq:hmix}, there is of course no such difference for the input 
mass.

\item
The variable \Code{\$InoScheme} (set \emph{before} model initialization)
determines the renormalization scheme for the chargino/neutralino 
sector.  Choices are \Code{CCN[$n$]} and \Code{CNN[$c$,$n$,$n'$]}; 
details in \refse{sec:cnscheme}.  The default is \Code{CCN[1]}.

\item

The variable \Code{\$SfScheme[\Vt,\,\Vg]} determines the renormalization 
scheme for sfermion $\Sf_{tg}$, $t = 4$ for squarks, $t = 2$ for 
sleptons of generation $g$.  Mathematica patterns may be used for \Vt\ 
and \Vg\ to combine definitions, \eg \Code{\$SfScheme[2,\,\uscore]}.  
Choices are \Code{DR[\Vi]} (for squarks) and \Code{OS[\Vi]} (for squarks 
and sleptons), where $i = 1, 2$ identifies the on-shell sfermion; 
details in \refeqs{eq:sqscheme} and \eqref{eq:slscheme}.  The default is 
\Code{DR[2]} for squarks and \Code{OS[2]} for sleptons.

\item
CKM mixing can be turned on with \Code{\$CKM = True} (set \emph{before} 
model initialization).  Be aware that in the present (minimally 
flavor-violating) version of the model file this may lead to incomplete 
cancellation of divergences if squarks, in particular external ones, are 
involved, see the discussion after \refeq{eq:dCKM}.
\end{itemize}
The model file prints out the settings of all relevant flags during 
initialization.


\subsection{Algebraic simplification}
\label{sec:algsimp}

Amplitudes generated with \Code{MSSMCT.mod} can directly be simplified 
with FormCalc.  Since the MSSM contains couplings that are relatively 
involved compared to QCD or the electroweak SM, it may become necessary 
to relax (some of) the fairly aggressive simplification functions to 
complete the algebraic simplification within reasonable run-time.

Potentially time-consuming transformations in the FORM part of the
calculation can be suppressed with the \Code{CalcFeynAmp} option 
\Code{NoCostly\,$\to$\,True}.

Upon return from FORM, FormCalc wraps a zoo of simplification functions 
around various parts of the amplitude.  All of these are `transparent' 
in the sense that they can be replaced by \Code{Identity} without 
affecting the numerical result.  The important ones are listed below, a 
complete inventory is given in the FormCalc manual.
\begin{itemize}
\item \Code{FormSub} is applied to subexpressions of an amplitude.
\item \Code{FormDot} is applied to combinations of dot products in an 
  amplitude.
\item \Code{FormMat} is applied to the coefficients of matrix elements 
  in an amplitude.
\item \Code{RCSub} works like \Code{FormSub} but is applied to 
  subexpressions of a renormalization constant.
\item \Code{RCInt} works like \Code{FormMat} but is applied to 
  coefficients of loop integrals in renormalization constants.
\end{itemize}


\subsection{Numerical Evaluation}
\label{sec:numeval}

In the FormCalc framework, the Mathematica expressions resulting from 
the algebraic simplification are translated to Fortran or C code for 
numerical evaluation.  The generated code has to be provided with the 
proper numerical values for the parameters appearing in the model, \ie 
the variables in \refta{tab:modelsyms}, derived from a (reasonably 
small) set of input parameters.  This is solved by a subroutine which is 
called at the beginning of the calculation to initialize all model 
parameters.

There are two ways to set up the MSSM parameters numerically:
\begin{itemize}
\item
The FormCalc module \Code{model\uscore mssm.F} provides stand-alone 
initialization of the MSSM parameters, \ie without linking to external 
libraries.  The user can choose to link with FeynHiggs 
\cite{feynhiggs,mhiggslong,mhiggsAEC,mhcMSSMlong}, however, to obtain 
the state-of-the-art Higgs-mass values (otherwise an approximation 
formula is used which includes the major two-loop shifts).  Details on 
the usage and the parameter relations employed in this approach are 
given in \citere{MSSMmod}.

\item
The module \Code{model\uscore fh.F} uses FeynHiggs as a frontend 
\cite{acat11,mhcMSSMlong}.  The code generated by FormCalc inherits thus 
the ability to read parameter files in either native FeynHiggs or SLHA 
format, and of course obtains all MSSM parameters and Higgs observables 
from FeynHiggs.  There is no duplication of initialization code this 
way, and moreover the parameters are consistent between the Higgs-mass 
calculation and the evaluation of the FormCalc-generated matrix 
elements.
\end{itemize}


\section{Tests}
\label{sec:tests}

A model of the complexity of the MSSM needs exhaustive checks to be 
trustable in all sectors.  For the tree-level couplings we could rely on 
the testing and maturity of the original \Code{MSSM.mod} model file 
\cite{MSSMmod}.  The new parts, \ie the counterterm vertices and the 
renormalization constants, have been tested in two ways.
\begin{itemize}
\item
On a relatively technical level, we ascertained numerically (and 
sometimes analytically) for a significant number of processes,
listed in \refta{tab:finite}, that the renormalization actually works, 
\ie that the results are UV- and IR-finite.

A version of \Code{MSSMCT.mod} with minor modifications (slightly 
different sbottom- and stau-sector renormalization, though with 
identical UV-finiteness properties, as well as finite mass shifts) had 
already been used to generate UV- and IR-finite heavy stop, sbottom, and 
stau decays \cite{Stop2decay,SbotRen,Stau2decay}, chargino and 
neutralino decays \cite{LHCxC,LHCxN}, and gluino decays 
\cite{Gluinodecay}, which are not listed in \refta{tab:finite}.

\begin{table}[p]
\caption{\label{tab:finite}Finite $1\to 2$ decays and $2\to 2$ 
scattering reactions computed with \Code{MSSMCT.mod}.  For the decays 
into photons, UV-finiteness was verified analytically.  $H^-\to G G^-$,
$H^-\to W Z$, and $H^-\to G^- Z$ are loop-induced processes.  Processes 
with external Goldstone bosons are unphysical and not necessarily
IR-finite.}
$$
\begin{array}{|l|}
\hline
h\to\{A A,\ H H,\ G G,\ G^+ G^-\} \\
h\to\{A\gamma,\ G\gamma,\ G Z,\ G^+ W^-\} \\
h\to \bar c c \\
\hline
H\to\{h h,\ A A,\ A G,\ G G,\ H^+ H^-,\ G^+ G^-\} \\
H\to\{A\gamma,\ G\gamma,\ A Z,\ G Z\} \\
H\to\{Z Z,\ W W\} \\
H\to \bar c c \\
\hline
A\to\{h G,\ G^+ G^-\} \\
A\to\{h\gamma,\ G\gamma,\ h Z,\ G^- W^+\} \\
A\to\{e^+ e^-,\ \bar u u\} \\
\hline
H^-\to\{h G^-,\ H G^-,\ A G^-,\ G G^-\} \\
H^-\to\{G^- Z,\ h W^-\} \\
H^-\to W^- Z \\
H^-\to\{\bar u d,\ \neu1\cham1\} \\
\hline\hline
h H\to A G \\
h A\to H G \\
\{h A,\ h G,\ H A,\ H G\} \to H^+ G^- \\
A A\to\{A A,\ G G,\ H^+ H^-\} \\
A G\to\{H^+ H^-, G^- H^+\} \\
G G\to\{G G,\ H^- G^+,\ G^- G^+,\ H^+ H^-\} \\
\{H^- H^+,\ H^- G^+\}\to H^- G^+ \\
\hline
h Z\to H^+ W^- \\
H H^-\to\{\gamma W^-,\ Z W^-\} \\
A Z\to H^+ W^- \\
A H^-\to Z W^- \\
\hline
\vrule width 0pt depth 0pt height 2.5ex
\{A A,\ G G,\ \bar\Stop_2\Stop_2\}\to \bar\Stop_1\Stop_1 \\
\{H H^-,\ A H^-\}\to \bar\Stop_1\Sbot_1 \\
\{A H^-,\ A G^-\}\to \Snu_\tau\Stau_1 \\
\hline
e^+ e^-\to\{Z h,\ W^+ H^-\} \\
\hline
\end{array}
$$
\end{table}

\item

The `gold standard' are tuned comparisons of selected reactions, 
however.  They are quite a bit more thorough but require a lot of work.
For lack of literature results to compare with, these comparisons were
mostly restricted to the MSSM with real parameters.

We calculated the decay $\Sbot_{1,2} \to \Stop_1 H^{-}$ and found good 
agreement with \citere{stopsbot.phi.als}.  We checked that we are in 
good agreement with \citere{glsqq.als} using their input parameters, 
where a small difference remains due to the different renormalization 
schemes.  $H\to h h$ is in perfect agreement with \citere{karina}, for 
both real and complex parameters, and $e^+e^-\to t\bar t$ agrees to 11 
digits with \citere{HoS99}.

We performed a detailed comparison with \citere{liebler} for the decay 
$\chapm{2}\to\neu{1} W^\pm$, where the chargino/neutralino sector is 
renormalized differently than in our prescription.  After a correction 
of the charge renormalization in \citere{liebler} we found good 
agreement at the level expected for different renormalization schemes in 
the chargino/neutralino sector.

We worked with the program \Code{SFOLD} \cite{sfold} to obtain numerical 
results for stau decays.  \Code{SFOLD} is restricted to the MSSM with 
real parameters and uses \DRbar\ renormalization throughout.  Although 
OS masses can be substituted on internal and/or external lines in 
\Code{SFOLD}, we preferred to use \DRbar\ masses in our calculation for 
the comparison.  \Code{SFOLD} also adopts a running electromagnetic 
coupling $\alpha_{\text{em}}(Q^{\DRbar})$ with a numerical value 
significantly higher than the $\alpha_{\text{em}}(0)$ used in our 
renormalization scheme (see \refse{sec:gauge}) and thus our tree-level 
results differ substantially.  At the one-loop level, the two results 
are in better agreement than expected, however.  This agreement improves 
for lower values of $Q$, but differences at the level of $5\%$ were 
found for $Q\sim 2$ TeV.

A comparison of different renormalization schemes in the 
chargino/neutralino sector can be found in \citere{LHCxN}.  The 
differences found with respect to another on-shell renormalization in 
the chargino/neutralino sector were small and of the expected size of 
two-loop contributions.
\end{itemize}


\section{Availability}

Starting from Version 3.9, the \Code{MSSMCT.mod} model file is included 
in the FeynArts distribution.  The package can be downloaded from 
\Code{http://feynarts.de}.

The FormCalc features described in \refses{sec:algsimp} and 
\ref{sec:numeval} are available from Version 8.2 on, which can be 
obtained from \Code{http://feynarts.de/formcalc}.

FeynArts and FormCalc each include a comprehensive manual which explains 
installation and usage.  Both are open-source programs and licensed 
under the LGPL.


\section*{Acknowledgements}

We thank G.~Weiglein for helpful discussions and P.~Drechsel and 
S.~Pa{\ss}ehr for bugfixes.  S.H.\ was supported by CICYT (Grant No.\ 
FPA 2010--22163-C02-01).  F.P.\ was supported by SEIDI and CPAN (Spain).  
S.H.\ and F.P.\ were also supported by the Spanish MICINN's 
Consolider-Ingenio 2010 Program under Grant MultiDark No.\ 
CSD2009-00064.


\newcommand\jnl[1]{\textit{\frenchspacing #1}}
\newcommand\vol[1]{\textbf{#1}}

\end{document}